\documentclass[%
 reprint,
 amsmath,amssymb,
 aps,
pra,
longbibliography]{revtex4-2}
\usepackage{bbold}
\usepackage{bm}
\usepackage{braket}
\usepackage[pagewise]{lineno}
\usepackage{csquotes}
\usepackage{derivative}
\usepackage{dcolumn}
\usepackage{dsfont}
\usepackage{float}
\usepackage{graphicx}    
\usepackage{lineno}
\usepackage{mathtools}
\usepackage{physics}
\usepackage{slashed}
\usepackage{tikz}
\usepackage{verbatim}
\usepackage{xcolor}

\usepackage{hyperref}

\usepackage{booktabs}
\usepackage{nicematrix} 
\definecolor{giallo}{RGB}{255, 255, 0}
\definecolor{rosso}{RGB}{255, 0, 0}
\definecolor{verde}{RGB}{0, 128, 0}
\definecolor{blu}{RGB}{0, 0, 255}
\definecolor{white}{RGB}{255, 255, 255}

\usepackage{float}

\begin{document}

\title{Local and Global Master Equations through the Lens of Non-Hermitian Physics}

\author{G. Di Bello$^{1,2,*}$}\author{F. Pavan$^{1}$}\author{V. Cataudella$^{2,3}$}\author{D. Farina$^{1,2,*}$} 
\affiliation{$^{1}$Dip. di Fisica E. Pancini - Università di Napoli Federico II - I-80126 Napoli, Italy}
\affiliation{$^{2}$INFN, Sezione di Napoli - Complesso Universitario di Monte S. Angelo - I-80126 Napoli, Italy}
\affiliation{$^{3}$SPIN-CNR and Dip. di Fisica E. Pancini - Università di Napoli Federico II - I-80126 Napoli, Italy}

\affiliation{$^*$Corresponding authors: G. Di Bello, grazia.dibello@unina.it and D. Farina, donato.farina@unina.it}

\date{\today}

\begin{abstract}
We investigate the relation between non-Hermitian Hamiltonian and Lindblad dynamics in nonequilibrium open quantum systems. Non-Hermitian models can extend phase diagrams and enable sensing advantages, but such effects often rely on postselection, raising questions about their relevance for unconditional dynamics. Using a minimal two-qubit setup mediating a heat current, we compare local and global Markovian master equations with their non-Hermitian counterparts. 
We observe that exceptional points emerge only in the local master equation and in the corresponding non-Hermitian Hamiltonian at sufficiently strong nonequilibrium. We further consider hybrid configurations, where one bath is treated with a Lindblad description and the other with a non-Hermitian approach, interpolating between the two extremes. 
Our results contribute understanding the role of quantum jumps and exceptional points in nonequilibrium open quantum systems and identify a simple, experimentally accessible architecture—realizable, for instance, in circuit-QED platforms—for their exploration.
\end{abstract}

\maketitle

\section{Introduction}

A detailed understanding of {open-system} dynamics is essential for addressing foundational questions and enabling progress in quantum technologies.
A central theoretical notion is the microscopic derivation of quantum master equations from an exact system–environment Hamiltonian.
The conventional route relies on the Born–Markov–{Secular} approximations \cite{breuer2007theory}.
In particular, the secular approximation—essentially a rotating-wave approximation—has recently been the focus of extensive investigations.
When applied, it leads to the global master equation, which is known to preserve thermodynamic consistency.
However, in regimes where the internal system couplings are small, the secular approximation is not necessarily reliable \cite{hofer2017markovian, gonzalez2017testing}.
In such cases, one may resort to local master equations (bypassing the secular approximation \cite{scali2021local}) which, although often more accurate in capturing the dynamics, must be used with care because they can violate the second law of thermodynamics \cite{levy2014local}; however, reconciliation proposals do exist \cite{de2018reconciliation, hewgill2021quantum}.
Local master equations are also widely employed in the modeling of open many-body systems and quantum simulations \cite{verstraete2009quantum,chen2025efficient}.
A sustained {debate} \cite{hofer2017markovian, gonzalez2017testing} has emerged concerning which form of the Lindblad master equation most faithfully captures the exact dynamics of multipartite systems.
Most investigations have concentrated on qubit and harmonic-oscillator architectures, owing to their analytical and computational tractability.
A number of works have compared local and global Lindblad master equations with exact dynamics, assessing their respective regimes of validity.
Prominent examples include analyses of nonequilibrium steady states (NESS) \cite{hofer2017markovian, gonzalez2017testing} and studies examining both steady-state and transient behavior in thermalization processes
of Gaussian bosonic systems
\cite{farina2020going}.
Furthermore, analytical treatments \cite{cattaneo2019local, khandelwal2020critical} have characterized steady-state properties in both approaches, identifying conditions under which the resulting steady state can exhibit entanglement. 
Comparing local and global master equations remains an {active topic} of investigation from several perspectives, including the statistical characterization of entanglement degradation \cite{9dw1-y4bl} and the Gaussian dynamics of harmonic chains \cite{babakan2025open1,babakan2025open2}.

It is well established that the Lindblad master equation can be {unraveled} in multiple ways, with photo-detection and homodyne detection serving as standard examples \cite{albarelli2024pedagogical, PhysRevA.106.042609}.
Among these, the non-Hermitian trajectory—which arises naturally in photo-detection–based unravelings—corresponds to the no-jump condition \cite{roccati2022non, guo2025designing}, and its occurrence probability generally decays exponentially with time.
{Non-Hermitian physics} \cite{ashida2020non} offers an effective framework for describing loss, gain, and measurement backaction, thereby extending conventional quantum dynamics in a physically meaningful way.
It also unveils peculiar phenomena—such as exceptional points \cite{heiss2012physics, minganti2019quantum, naghiloo2019quantum,khandelwal2021signatures,svegborn2026framework} and non-reciprocal responses \cite{PhysRevResearch.2.013058, RevModPhys.93.015005}.

A systematic comparison between Lindblad and non-Hermitian Hamiltonian dynamics is therefore warranted, both from a foundational standpoint \cite{minganti2019quantum} and for translating the topological features and sensing advantages of non-Hermitian Hamiltonian models into the Lindblad setting, where dynamics require no postselection and steady states are well defined \cite{PhysRevX.9.041015,mcdonald2020exponentially}.
These questions are also actively explored in many-body contexts \cite{chaduteau2025lindbladian}, and novel hybrid configurations interpolating between Lindblad and non-Hermitian Hamiltonian scenarios \cite{gu2025exploring, settimo2025stochastic}. 

Motivated by these considerations, it is timely to examine local and global master equations through the lens of non-Hermitian physics.
{In this work}, we investigate the connection between non-Hermitian Hamiltonian dynamics and Lindblad dynamics in nonequilibrium open quantum systems.
Using a minimal two-qubit architecture that mediates a heat current, we compare local and global Lindblad master equations with their non-Hermitian counterparts, obtained by dropping the quantum jump terms.
We show that exceptional points arise in the local Lindblad master equation and in the associated non-Hermitian no-jump trajectory when the system is driven sufficiently far from equilibrium.
In contrast, we observe that global master equations are typically inadequate for capturing exceptional points.
We further examine hybrid configurations in which one bath is modeled with a Lindblad description and the other with a non-Hermitian approach, thereby interpolating between the two limiting regimes.
Our results uncover novel regimes for exceptional-point engineering and deepen the understanding of non-Hermitian phenomena in nonequilibrium settings.

{The paper is organized as follows. 
In Sec.\,\ref{sec:prel} we introduce the necessary preliminary concepts, including selected aspects of Lindblad and non-Hermitian evolutions, exceptional points, and steady states.
We present the paradigmatic local and global master equations for a pair of interacting qubits, each coupled to a thermal bath at a different temperature.
In Sec.\,\ref{sec:results} we detail the main results of this work, briefly outlined above.
Finally, in Sec.\,\ref{sec:conclusions}, we summarize the main findings and discuss possible future directions for this line of research.
}

\section{Preliminary concepts}
\label{sec:prel}
\subsection{Lindblad and non-Hermitian Hamiltonian dynamics}
Non-Hermitian Hamiltonians can be derived from the Lindblad master equation using two different physical interpretations: a semiclassical approximation and postselection through quantum trajectory unraveling \cite{minganti2019quantum,albarelli2024pedagogical}.
Following Ref.\,\cite{minganti2019quantum},
in the semiclassical approximation, the effective non-Hermitian Hamiltonian is obtained by neglecting the effect of quantum jumps. 
The term ``jump'' originates from the quantum trajectory picture, where it refers to the abrupt stochastic change of the wave function due to a discrete interaction with the environment. By disregarding these jumps, the system’s evolution is effectively governed by a non-Hermitian Hamiltonian term that incorporates a continuous nonunitary dissipation term.
In contrast, in the postselection approach, the system’s dynamics are described within the quantum trajectory framework, where the environment is continuously monitored. In this formalism, the system state is represented by a stochastic wave function that evolves smoothly according to the effective non-Hermitian Hamiltonian in a single trajectory when no quantum jumps occur. This approach provides a physically rigorous interpretation of non-Hermitian Hamiltonian dynamics as the conditional evolution of the system, given that no quantum jumps have been detected.

The time evolution of the reduced density matrix $\rho_L(t)$ of an open quantum system described by the Lindblad master equation $\dot{\rho}_L=\mathcal{L}\rho_L$, with $\mathcal{L}$ the Lindblad superoperator generating the dynamics, has the following form:

\begin{eqnarray}\label{eq:rholind}
&&\dot{\rho}_L(t) = -i\left[\mathcal{H},\rho_L(t)\right] +\sum_{j\geq 1}\kappa_j
\mathcal{D}\left[L_j\right]\rho_L(t)
\end{eqnarray}
where the subscript $L$ indicates that the evolution follows the Lindblad formalism, $\kappa_j$ are the relaxation rates, and $L_j$ are the Lindblad operators effectively describing the system-environment interaction. The first term represents the coherent evolution, while the summation term includes the nonunitary dissipative contributions, represented by the dissipators $\mathcal{D}\left[L_j\right]$ acting on the reduced density matrix as $\mathcal{D}\left[L_j\right]\rho_L(t):=L_j\rho_L(t)L_j^{\dagger}-\frac{1}{2}\{L_j^{\dagger}L_j,\rho_L(t)\}$. 
The GKSL equation in Eq.\,\eqref{eq:rholind} can be cast 
as \cite{minganti2019quantum}
\begin{equation}
\dot{\rho}_L(t)
= -i\left(\mathcal{H}_{\text{eff}}\rho_L(t)-\rho_L(t)\mathcal{H}_{\text{eff}}^{\dagger}\right) + \sum_{j\geq 1}\kappa_j L_j \rho_L(t)L_j^{\dagger},
\end{equation}
where we introduced the effective non-Hermitian Hamiltonian,
\begin{equation}\label{eq:nheq}
\mathcal{H}_{\text{eff}}
=
\mathcal{H} - i \Gamma\,,
\quad
\Gamma= \frac{1}{2} \sum_{j\geq 1} \kappa_j L_j^{\dagger}L_j\,,
\end{equation}
with $\mathcal{H}$ and $\Gamma$ being Hermitian operators.
We also notice that $\Gamma$ is a positive semidefinite operator, because the $\kappa_j$'s are positive.
Specifically, $\mathcal{H}$ captures the coherent evolution and $\Gamma$ includes the dissipative effects due to the environment. The second sum in Eq.\,\eqref{eq:rholind} accounts for the quantum jumps which are neglected in a semiclassical approximation, thereby recovering a purely non-Hermitian Hamiltonian evolution \cite{sim2025oservables}.

The same non-Hermitian Hamiltonian evolution can be formally obtained via postselection. Following, e.g., Refs.\,\cite{minganti2019quantum, guo2025designing}, 
one can consider the Kraus operators which define the quantum trajectories necessary for the unraveling process \cite{guo2025designing}, excluding those that describe the jumps. A possible form for the Kraus operators $K_0$ and $K_{j\geq 1}$ is given by,
\begin{align}
K_0 &= \mathbb{1} - i\,\mathcal{H}_{\rm eff}\, dt\,, \\
K_j &= L_j\sqrt{\kappa_j dt}\,, \quad \text{for } j\geq 1\,.
\end{align}
These operators describe the stochastic evolution of the system, where $K_0$ corresponds to the no-jump deterministic evolution governed by $\mathcal{H}_{\text{eff}}$, while the $K_j$'s describe the discrete jumps induced by the environment.
By dropping the quantum jump terms, the master equation is no longer trace preserving, leading to an unnormalized density operator. A physical state can then be recovered by normalizing with its trace, which corresponds to the evolution conditioned on the no-jump process. This defines the state under non-Hermitian Hamiltonian dynamics.
The normalized state can also be obtained via the non-linear master equation \cite{sergi2016quantum},
\begin{equation}
\label{NHH-nonlinearME}
\dot{\rho}_{\rm nH}(t)=-i[\mathcal{H},\rho_{\rm nH}(t)]-\{\Gamma,\rho_{\rm nH}(t)\}+2\rho_{\rm nH}\text{Tr}\left( \Gamma \rho_{\rm nH}(t) \right).
\end{equation}

\subsection{Steady states}
The steady state of a Lindblad evolution is given by a density operator that is a fixed point of the Lindbladian, i.e., an eigenoperator with zero eigenvalue and unit trace. At least one such state always exists.
In the non-Hermitian Hamiltonian framework there exist multiple states that can persist indefinitely. Specifically, if the system is initially prepared in a right eigenvector of the non-Hermitian Hamiltonian, it will remain in that state. 
However, let us consider a “naturally” prepared initial state—which inevitably includes some level of noise—having a nonzero (even arbitrarily small) overlap with all right eigenvectors.
We define the decay rates as minus the imaginary parts of the eigenvalues of $\mathcal{H}_{\rm eff}$. 
Let us consider the minimum decay rate and assume that it is nondegenerate.
The system will asymptotically evolve towards a $\ket{\psi_{\min}}$, corresponding to the minimum decay rate.
This is the longest-lived right eigenvector. 
The degenerate case follows straightforwardly.

\subsection{Local and global master equations}\label{sec:loc-glob}
\begin{figure}[ht]
    \centering
    \includegraphics[width=0.9\linewidth]{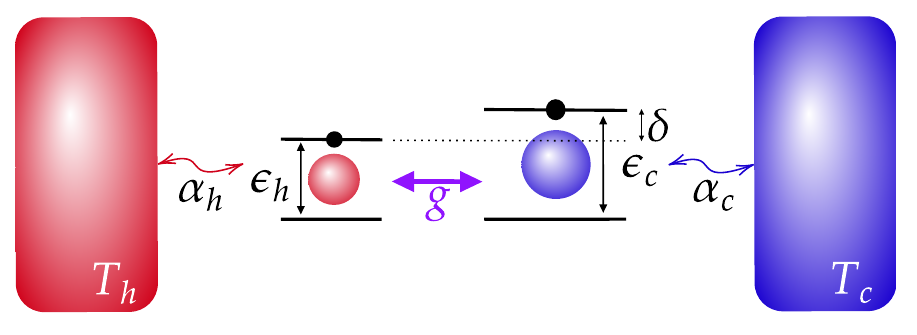}
    \caption{Schematic representation of the model studied in this work. Two detuned qubits interacting with each other and coupled to independent hot and cold bosonic baths.}
    \label{fig:model}
\end{figure}

We now introduce the Lindblad generators considered in this work.
We focus on the dissipative model shown in Fig.\,\ref{fig:model}, consisting of two detuned qubits ($h$,$c$) coupled with strength $g$ via a $\sigma_x^h\sigma_x^c$ interaction. 
The Hamiltonian of the open system reads
\begin{equation}
\mathcal{H}=\epsilon_h \sigma_+^h \sigma_-^h
+
\epsilon_c \sigma_+^c \sigma_-^c
+
g \sigma_x^h\sigma_x^c\,.
\label{ham}
\end{equation}
Each qubit is coupled to an independent bosonic bath at a different temperature 
(however, similar reasoning can be analogously applied to fermionic baths \cite{PhysRevB.97.085435,PhysRevResearch.2.033497,PhysRevA.100.042327}).
The hot qubit sets the system’s energy scale $\epsilon_h$, with the detuning being $\delta=\epsilon_c-\epsilon_h$. 
For the local approach master equation, the evolution of the reduced density matrix ${\rho}_L^l(t)$ (where the superscript $l$ stands for local) is described by a Lindbladian in the form of Eq.\,\eqref{eq:rholind}, with $\mathcal{H}$ in Eq.\,\eqref{ham} the Hamiltonian of the two detuned interacting qubits. The nonunitary part takes the form, 
\begin{equation}\label{eq:dissipl}
\sum_{j\in\{h,c\}}\left(\gamma_j^+ \mathcal{D}\left[\sigma_+^{(j)}\right]\rho_L^l(t)+\gamma_j^- \mathcal{D}\left[\sigma_-^{(j)}\right]\rho_L^l(t)\right),
\end{equation}
where there are four dissipators $\mathcal{D}\left[\sigma_{\pm}^{(j)}\right]$, and we introduce
\begin{align}
\gamma_j^+ &= \gamma_j(\epsilon_j) n_B^j(\epsilon_j),\\
\gamma_j^- &= \gamma_j(\epsilon_j) (1+n_B^j(\epsilon_j)),
\nonumber
\end{align}
with $\gamma_j(\epsilon_j)$ being the spontaneous emission decay rates, related to the bath spectral density via
\begin{equation}
    \gamma_j(\omega) = 2\pi J_j(\omega).
\end{equation}
We choose Ohmic spectral densities of the form $J_{j}(\omega) = \frac{\alpha_j}{2}\omega \Theta(\omega_c-\omega)$, where $\alpha_j$ controls the system-bath coupling (see Fig.\,\ref{fig:model}), $\Theta$ is the Heaviside function and $\omega_c$ is the cutoff frequency, the largest energy scale in the system. Moreover, $n_B^j(\epsilon_j)$ denotes the Bose-Einstein distribution for the $j$-bath, evaluated at energy $\epsilon_j$.
On the other hand, for the global master equation, the evolution of the reduced density matrix ${\rho}_L^g(t)$ (where the superscript $g$ stands for global) is described by another Lindbladian with the same Hamiltonian $\mathcal{H}$ and with the nonunitary part given by
\begin{eqnarray}\label{eq:dissipg}
\sum_{\eta}\sum_{j\in\{h,c\}}\big(\gamma_j^+(\xi_{\eta}) \mathcal{D}[L_j^{\dagger}(\xi_{\eta})]\rho_L^g(t)+
\\\nonumber
\gamma_j^-(\xi_{\eta}) \mathcal{D}\left[L_j(\xi_{\eta})\right]\rho_L^g(t)\big),
\end{eqnarray}
where the dissipators now refer to the Lindblad operators $L_j(\xi_{\eta})$ and their adjoints, which are associated with the transitions between the eigenstates of the two-qubit Hamiltonian in Eq.\,\eqref{ham}, corresponding to the four eigenvalues $\xi_{\eta}$ of the two-qubit Hamiltonian. 
The kinetic coefficients assume the expressions
\begin{align}
\gamma_j^+(\xi_{\eta}) &= \gamma_j (\xi_{\eta})n_B^j(\xi_{\eta}),\\
\gamma_j^-(\xi_{\eta}) &= \gamma_j (\xi_{\eta})(1+n_B^j(\xi_{\eta})),
\nonumber
\end{align}
with the only difference from the local approach being the energy at which these functions are evaluated. Here, the energy corresponds to the possible transition frequencies between the eigenstates of the two-qubit Hamiltonian, rather than the individual qubit energies $\epsilon_j$. For the explicit analytical forms of the jump operators in the Lindblad master equation for a related model, see Ref.\,\cite{cattaneo2019local}.

\subsection{Error metrics}
To compare different states we shall use the trace distance
\begin{equation}\label{eq:trdist}
D(\rho,\sigma) = \frac{1}{2} \left\Vert \rho - \sigma \right\Vert_1\,,
\end{equation}
with $\left\Vert A \right\Vert_1 := \text{Tr}|A|$,
a measure of distinguishability between two quantum states, $\rho$ and $\sigma$ \cite{helstrom1969quantum}.
This metric is zero if the two states are identical and one if they are orthogonal.
In our context, for instance, the trace distance can be used to verify convergence of the evolved state to the steady state (see Appendix \ref{app:trace_dist_ss}) and to compare the local and global approaches throughout the dynamics, as discussed below.
Furthermore, in the next section we will present numerical results dealing with quantum operations that are not necessarily trace preserving. 
Indeed, as previously mentioned, in the non-Hermitian Hamiltonian setting, the physical states are the postselected (no-jump) states obtained by a posteriori normalization.
To include these cases, we introduce the trace distance of the evolved states through different evolutions $\mathcal{E}_t$ and $\mathcal{F}_t$, from time $0$ to time $t$, not necessarily trace preserving,
\begin{equation}
\label{trace-distance-output-norm}
\frac{1}{2}
\left\Vert 
\frac{\mathcal{E}_t(\rho)}{{\rm Tr}[\mathcal{E}_t(\rho)]}
-
\frac{\mathcal{F}_t(\rho)}{{\rm Tr}[\mathcal{F}_t(\rho)]}
\right\Vert_1\,,
\end{equation}
$\rho$ being the initial state.
We shall also make use of a recently introduced metric \cite{shi2023error} that generalizes the concept of the diamond distance \cite{aharonov1998quantum,gilchrist2005distance} to the case of non-trace-preserving quantum operations.
Following Ref.\,\cite{shi2023error},
the \textit{general distance} is defined as
\begin{equation}
d_g(\mathcal{E}_t, \mathcal{F}_t):=\frac{1}{2}\max_\rho
\left\Vert 
\frac{(\mathcal{E}_t\otimes \mathcal{I})(\rho)}{{\rm Tr}[(\mathcal{E}_t\otimes \mathcal{I})(\rho)]}
-
\frac{(\mathcal{F}_t\otimes \mathcal{I})(\rho)}{{\rm Tr}[(\mathcal{F}_t\otimes \mathcal{I})(\rho)]}
\right\Vert_1,
\label{dg}
\end{equation}
where $\mathcal{I}$ represents the identity operation on an auxiliary space. Eq.\,\eqref{dg} consistently reduces to the diamond distance when the operations are trace preserving. 
The quantity $d_g$ quantifies the distance between the quantum operations themselves by maximizing over all input states defined on an extended space, thereby making the measure independent of the specific input.
Despite performing the optimization over the input state can be a difficult task, one can obtain an upper bound 
on this quantity that reads as follows
\begin{equation}\label{eq:upperbound}
d_g
(\mathcal{E}_t,\mathcal{F}_t) \leq d_g^{\rm (ub)}(\mathcal{E}_t,\mathcal{F}_t)\,,
\end{equation}
with $d_g^{\rm (ub)}(\mathcal{E}_t,\mathcal{F}_t):=d_\diamond(\mathcal{U}_t,\mathcal{V}_t) + d_g(\mathcal{I},\mathcal{M}_t)\,.$
Without entering too deeply into technicalities,
$d_\diamond(\mathcal{U}_t,\mathcal{V}_t)$ denotes the conventional diamond distance between two renormalized trace-preserving operations, while $d_g(\mathcal{I},\mathcal{M}_t)$ represents the general distance between the identity operation and the normalization operation $\mathcal{M}_t$, referred to as the normalization distance. 
While the bound $d_g^{\rm (ub)}$ might not always be tight, if the operations under comparison exhibit only
infinitesimal deviations from being trace-preserving, the gap
$d_g^{\rm (ub)}-d_g$ vanishes
(see Ref.\,\cite{shi2023error} for explicit expressions and further details).
%
Since Eq.\,\eqref{dg} defines $d_g$ as a maximization over input states $\rho$ on the enlarged space, evaluating the quantity for any fixed $\rho$ provides a lower bound on $d_g$. In particular, for a product input $\rho=\rho_S\otimes\rho_A$, with $\rho_A$ an arbitrary auxiliary state, Eq.\,\eqref{dg} reduces to Eq.\,\eqref{trace-distance-output-norm}, i.e., the trace distance between the corresponding normalized output states. Hence Eq.\,\eqref{trace-distance-output-norm} yields a specific lower bound on $d_g$.

\subsection{Non-Hermitian entropy production}
\label{sec:thermo-prel}
The quantification of entropy production in the Lindblad framework is well established \cite{kosloff2013quantum, vinjanampathy2016quantum}, while it is still debated in the non-Hermitian Hamiltonian setting. In this work we will adopt the proposal of Refs.\,\cite{sergi2016quantum,sergi2019density}, providing new use cases in the context of non-Hermitian Hamiltonian dynamics under nonequilibrium conditions induced by thermal bias.
The commonly used von Neumann entropy of the normalized state, $S_{\rm vN} = -\text{Tr}(\rho_{\rm nH} \ln \rho_{\rm nH})$, is not suitable for capturing the gain or loss of probability in the subsystem \cite{sergi2019density}. 
A quantity that is instead sensitive to this, is the non-Hermitian Hamiltonian von Neumann entropy \cite{sergi2016quantum,sergi2019density}, which is defined in terms of the non-normalized density matrix $\Omega_{\rm nH}$ as
\begin{equation}\label{eq:nHentr}
    S_{\rm nH} = -\text{Tr}\left( \rho_{\rm nH} \ln(\Omega_{\rm nH}) \right),
\end{equation}
where the units are in terms of both $\hbar$ and the Boltzmann constant $k_B$.
In contrast, the operator $\ln(\Omega_{\rm nH})$ tracks the evolution of the probability weight and leads to the following expression for the entropy production rate:
\begin{equation}\label{eq:nHentrprod}
    \dot{S}_{\rm nH} = 2 \left[ \, \text{Tr} \left( \Gamma \rho_{\rm nH} \ln(\Omega_{\rm nH}) \right) + \left( S_{\rm nH} + 1 \right) \, \text{Tr} \left( \Gamma \rho_{\rm nH}\right)\right],
\end{equation}
where each state depends on time $t$.

The difference between $ S_{\rm nH} $ and $ S_{\rm vN} $ quantifies how much $ \text{Tr}(\Omega_{\rm nH}) $ deviates from unity. Unlike $ S_{\rm vN} $, the quantity $ S_{\rm nH} $ is not invariant under constant imaginary shifts of the non-Hermitian Hamiltonian that preserve the form of the normalized density matrix $ \rho_{\rm nH} $. As we shall see explicitly later, this property allows $ S_{\rm nH} $ to capture the temperature dependence in non-Hermitian Hamiltonian dynamics.
The non-Hermitian Hamiltonian entropy can be interpreted as an analogue of the irreversible entropy production in Lindblad dynamics, with its time derivative corresponding to the entropy production rate, which can be rewritten as
\begin{equation}\label{eq:nHentrprod2}
\dot{S}_{\rm nH} = \dot{S}_{\rm vN} + 2\text{Tr}\left( \Gamma \rho_{\rm nH} \right).
\end{equation}
In case of Lindblad dynamics, we recall that the rate of irreversible entropy production $\dot{S}_{L}^i$ can be written as follows
\begin{equation}\label{eq:entrprodL}
    \dot{S}_{\rm L}^i=\dot{S}_{\rm vN}-\sum_{j\in\{h,c\}}\frac{\dot{Q_j}}{T_j},
\end{equation}
where $\dot{S}_{\rm vN}=-\text{Tr}\left( \mathcal{L} (\rho_{L}) \ln(\rho_{L}) \right)$ and $Q_j$ is the heat flowing into the system. 

\subsection{Exceptional points in non-Hermitian Hamiltonian and Lindblad dynamics}\label{subsec:eps_pre}
Exceptional points are a hallmark of non-Hermitian physics.
Let us consider a non-Hermitian Hamiltonian $H_{\rm eff}(\vec{\lambda})$ dependent on an array $\vec{\lambda}$ of classical parameters. Entries in $\vec{\lambda}$
can represent different quantities in parameter space, such as temperatures in the setup, Hamiltonian couplings, external fields, decay rates, etc.
Exceptional points are points in the parameter space where two (or more) eigenvalues and their corresponding eigenvectors coalesce \cite{minganti2019quantum}.
Concretely, this means that (i) the complex eigenvalues of the effective non-Hermitian Hamiltonian merge.
(ii) At the exceptional point, the right (left) eigenvectors of the effective Hamiltonian become linearly dependent, leading to a defective Jordan block structure. 
This can be quantified using the overlap measure $1 - |\langle v_i | v_j \rangle|^2$ for each pair of right eigenvectors $v_i,v_j$. Accordingly, at the exceptional point, this quantity vanishes for each pair that coalesces at the exceptional point.
Unlike in Hermitian systems, where eigenvalues remain distinct under small perturbations, non-Hermitian Hamiltonian systems can exhibit a qualitative change in their spectral properties at these critical points.
Another peculiar feature is
(iii) the divergence of the condition number of the right eigenvector matrix. 
The condition number of the matrix $V$, whose columns are the right eigenvectors of the non-Hermitian Hamiltonian, becomes large near an exceptional point \cite{belinschi2017squared,pick2019robust}.
The condition number is defined as
\begin{equation}
    \kappa(V) = \frac{\sigma_{\max}(V)}{\sigma_{\min}(V)},
\end{equation}
where $\sigma_{\max}(V)$ and $\sigma_{\min}(V)$ are the largest and smallest singular values of $V$, respectively. A large $\kappa(V)$ indicates that the eigenvectors are nearly linearly dependent, making the system highly sensitive to perturbations.

(iv)\,The associated dynamical consequences constitute another distinctive element. Near an exceptional point, the system exhibits enhanced sensitivity to perturbations, and the evolution becomes qualitatively different compared to regimes far from the exceptional point.

When the dynamics is governed by a Lindbladian, a Liouvillian exceptional point corresponds to a point in parameter space where two eigenmatrices of the Liouvillian coalesce \cite{minganti2019quantum}. This notion naturally extends the phenomenology of exceptional points from non-Hermitian Hamiltonians to open quantum systems described by dissipative dynamics.

\section{Results}\label{sec:results}
Having prepared the ground,
we present our results considering the nonequilibrium dissipative model depicted in Fig.\,\ref{fig:model}.
In Sec.\,\ref{subsec:localglobal} we compare the local and global 
master equations using a non-Hermitian Hamiltonian description.
We then relate the
Lindbladian dynamics to their corresponding non-Hermitian Hamiltonian dynamics.
We also discuss thermodynamic features, comparing the entropy production in the Lindblad and non-Hermitian Hamiltonian frameworks.
In Sec.\,\ref{subsec:NHH_EP} we demonstrate the presence (absence) of exceptional points in the non-Hermitian Hamiltonian local (global) master equation.
We extend the discussion to Lindblad dynamics, including hybrid configurations where the interaction with one bath is described by the non-Hermitian Hamiltonian term and the interaction with the other bath with the Lindblad dissipator.

\subsection{Comparing the different dynamics}\label{subsec:localglobal}
The non-Hermitian Hamiltonian evolution associated with the local master equation is obtained dropping the quantum jump terms in Eq.\,\eqref{eq:dissipl}.
It reads as
\begin{equation}\label{eq:nhl}
\dot{\Omega}_{\rm nH}^l(t)=-i\left(\mathcal{H}_{\text{eff}}^l\Omega_{\rm nH}^l(t)-\Omega_{\rm nH}^l(t)\mathcal{H}_{\text{eff}}^{l\,\dagger}\right),
\end{equation}
where
$\Omega_{\rm nH}$ is the non-normalized state and $\mathcal{H}_{\text{eff}}^l$ is the non-Hermitian Hamiltonian, derivable from Eq.\,\eqref{eq:nheq} using the appropriate local jump operators. 
Similarly, the global version of the non-Hermitian Hamiltonian evolution is governed by the equation,
\begin{equation}\label{eq:nhg}
\dot{\Omega}_{\rm nH}^g(t)=-i\left(\mathcal{H}_{\text{eff}}^g\Omega_{\rm nH}^g(t)-\Omega_{\rm nH}^g(t)\mathcal{H}_{\text{eff}}^{g\,\dagger}\right).
\end{equation}
In this work, the dynamical evolution is obtained by applying the non-Hermitian Hamiltonian Liouvillian to the state, represented as a vector in Liouville space \cite{gilchrist2009vectorization}.
Indeed, 
both Eqs.\,\eqref{eq:nhl} ($\alpha=l$) and \eqref{eq:nhg} ($\alpha=g$), can be expressed in the form
$\dot{\Omega}^{\alpha}_{\rm nH}(t)=\mathcal{L}'_{\alpha}({\Omega}_{\rm nH}(t))$. 
This defines the Liouvillian superoperator
$\mathcal{L}'_{\alpha}$
and we have ${\Omega}_{\rm nH}^{\alpha}(t)=e^{t \mathcal{L}'_{\alpha}}
{\rho}(0)$.

\begin{figure*}[ht]
    \centering
    \includegraphics[width=0.35\linewidth]{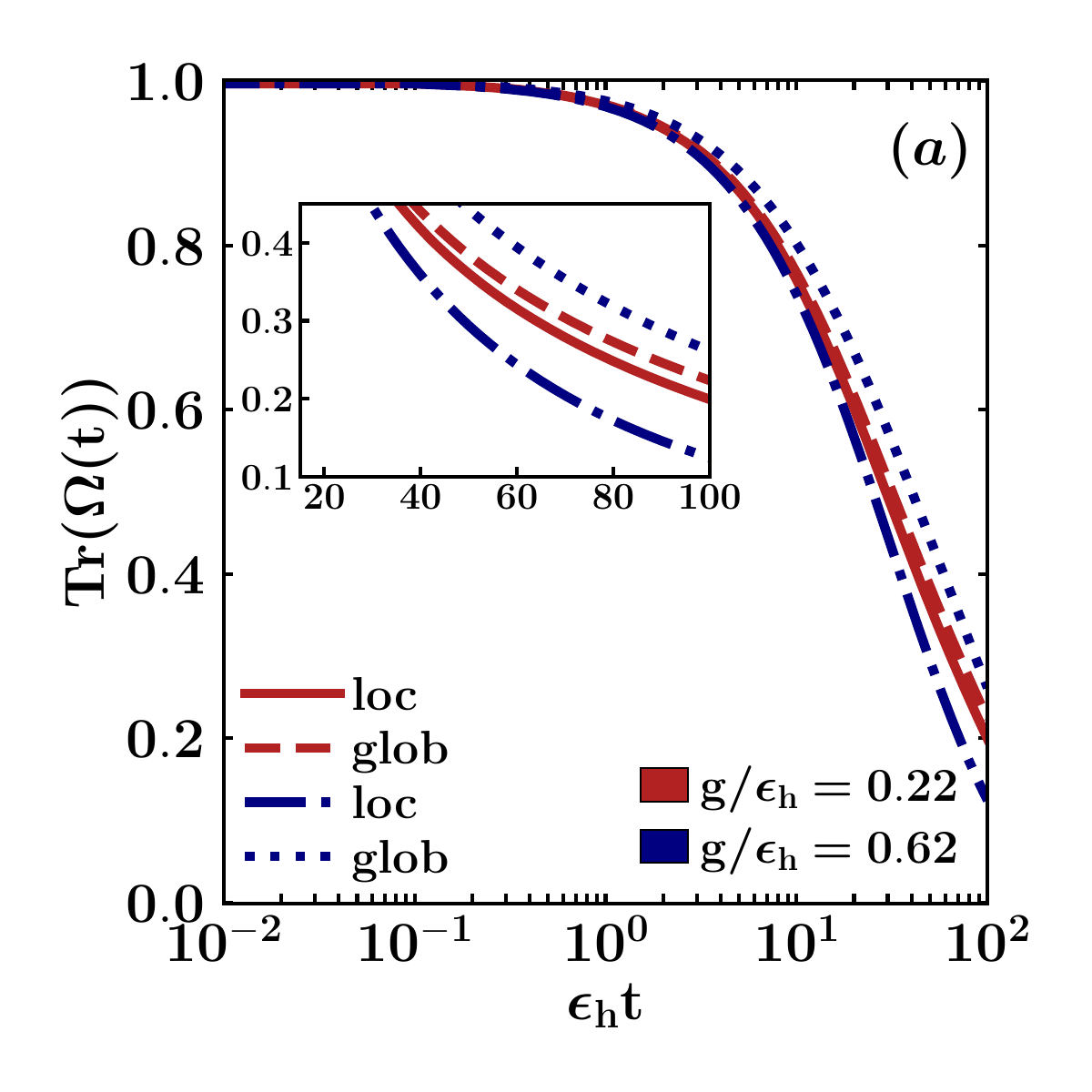}
    \includegraphics[width=0.35\linewidth]{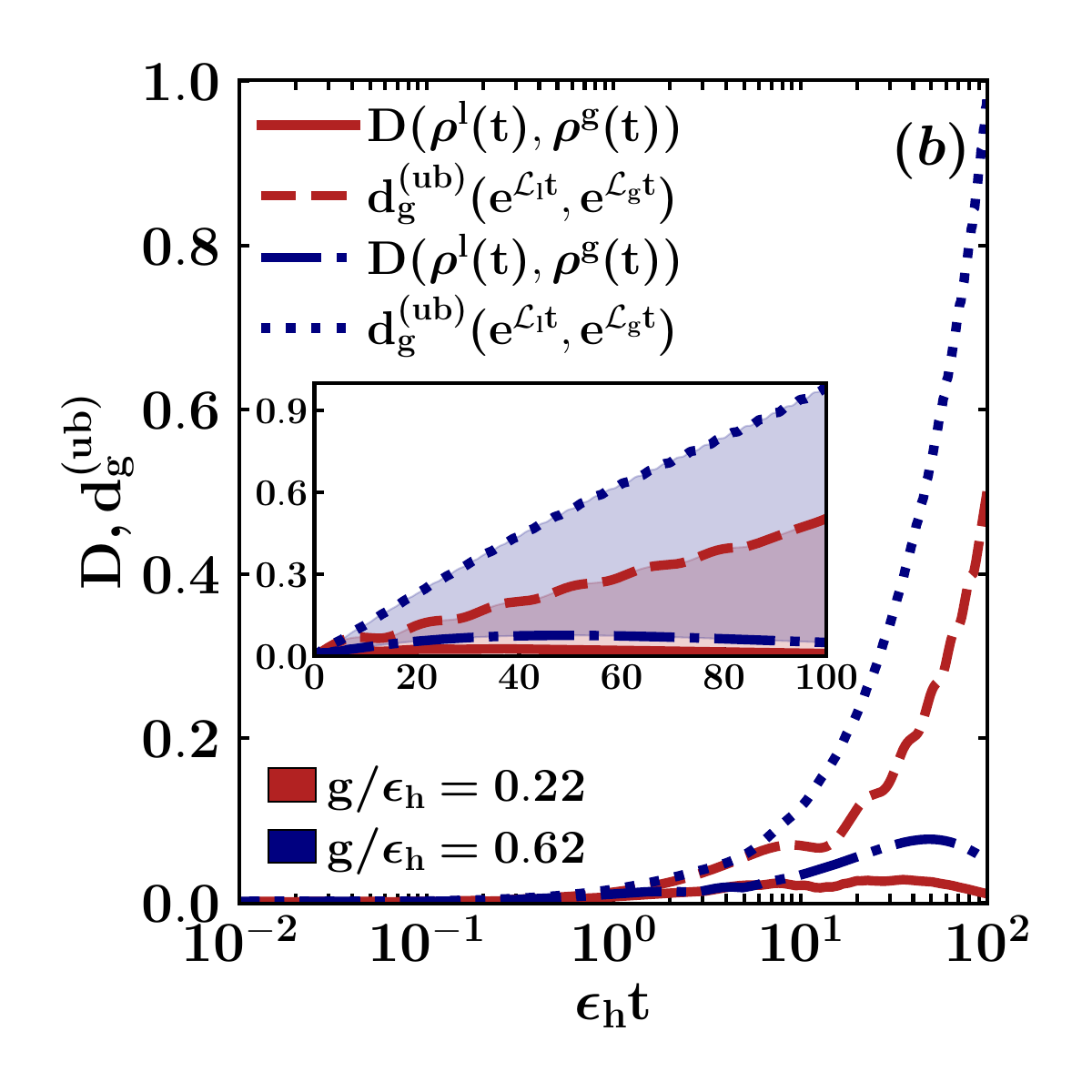}
    \caption{($a$)\,Evolution of the trace of the non-normalized state $\Omega_{\rm nH}$ as a function of dimensionless time for two values of the coupling $g$, plotted on a logarithmic scale, with a zoomed-in view shown in linear scale. ($b$)\,Trace distance in Eq.\,\eqref{eq:trdist} and upper bound in Eq.\,\eqref{eq:upperbound} as functions of dimensionless time, for the same two values of the coupling $g$, on a logarithmic scale. The inset provides a zoomed-in view with shaded regions highlighting the range between the upper bound and the trace distance (i.e., a lower bound), on a linear scale. The normalized non-Hermitian Hamiltonian evolved state $\rho_{\rm nH}(t)$ in the local (loc) approach (see Eq.\,\eqref{eq:nhl}) is compared with the global (glob) approach (see Eq.\,\eqref{eq:nhg}).
    The parameters used in our simulations are:
    $\epsilon_c = \epsilon_h$, $\alpha_c = 0.02\epsilon_h$, $\alpha_h = 0.005\epsilon_h$, $T_c = 0.1\epsilon_h$, $T_h = \epsilon_h$, $\omega_c = 10\epsilon_h$. We tune the coupling parameter $g \in \{0.22,0.62\} \epsilon_h$ (see legend).
        }
    \label{fig:lg}
\end{figure*}

The nonunitary evolution generated by the non-Hermitian Hamiltonian leads to a non-conserved trace, whose value has a direct probabilistic interpretation as the survival probability of the no-jump trajectory up to time $t$.
In this regard,
Fig.\,\ref{fig:lg}($a$) shows the time evolution of ${\rm Tr}(\Omega_{\rm nH})$, both for local and global approaches and for two values of the coupling $g$. 
From inset of Fig.\,\ref{fig:lg}($a$), we observe that the trace decays over time for both approaches, with behavior strongly dependent on the coupling strength $g$. Specifically, in the local approach, the trace decays faster as $g$ increases, while in the global approach, the opposite trend is observed: the decay slows down with increasing $g$. As a result, the difference between local and global dynamics becomes more pronounced at long times when comparing the dynamics for different $g$.
However, in the considered time window, the probability of observing non-Hermitian Hamiltonian trajectories remains non-negligible.
The (normalized) density matrix of the system is a posteriori obtained as
$\rho_{\rm nH}(t) := \Omega_{\rm nH}(t) / \text{Tr}[\Omega_{\rm nH}(t)]$.
\subsubsection{Local vs. global using non-Hermitian Hamiltonian}
We now turn to a comparison between the local and global approaches using non-Hermitian Hamiltonian dynamics, which, in contrast to the Lindblad case, remain largely unexplored in the literature. 
In the following numerical results, we fix the initial state as a thermal state with temperature $T_h$ and Hamiltonian $\mathcal{H}$ (defined in Eq.\,\eqref{ham}),
\begin{equation}
\label{thermal-state}
    \rho(0):=\frac{e^{-\mathcal{H}/T_h}}
    {{\rm Tr}(e^{-\mathcal{H}/T_h})}\,.
\end{equation}
This state can be experimentally obtained by first letting the system thermalize in contact with the hot bath and then connecting it to the cold bath to obtain the desired dynamics.
In Fig.\,\ref{fig:lg}($b$) we compare the non-Hermitian Hamiltonian local and global evolutions using the trace distance $D(\rho^l_{\rm nH}(t), \rho^g_{\rm nH}(t))$ (see Eqs.\,\eqref{eq:trdist} and \eqref{trace-distance-output-norm}).
We also report
the upper bound on the general distance between the corresponding non-trace-preserving quantum operations.
The quantum operations, $e^{t \mathcal{L}'_l}$ and $e^{t \mathcal{L}'_g}$, are
defined by the Liouvillians 
$\mathcal{L}'_l$ and $\mathcal{L}'_g$ previously introduced and the upper bound that we use is $d_g^{\rm(ub)}(e^{t \mathcal{L}'_l}, e^{t \mathcal{L}'_g})$ (see Eq.\,\eqref{eq:upperbound}). 
We plot this bound over the same time window and for the same values of $g$.
We observe that the upper bound remains below 1 throughout the entire time window, indicating that it provides meaningful information. 
As mentioned in the previous section, the trace distance serves, instead, as a lower bound on the general distance.
Accordingly, in the inset, we plot the shaded region between the trace distance and the generalized upper bound for the two selected values of $g$, illustrating the range within which the general distance lies. 
In the considered time window, the trace-distance analysis shows that the local and global non-Hermitian Hamiltonian trajectories become clearly distinguishable for times $t\gtrsim 1/\epsilon_h$.

\subsubsection{Lindblad vs. non-Hermitian Hamiltonian}
{
In Fig.\,\ref{fig:LnH},
we numerically compare Lindblad and non-Hermitian Hamiltonian dynamics, separately for the local approach (Fig.\,\ref{fig:LnH}($a$)) and the global approach (Fig.\,\ref{fig:LnH}($b$)). 
This is done through the trace distance between the evolved (normalized) states, that is shown as a function of time for four values of the coupling strength $g$. 
As one can expect, we observe that the two states are nearly indistinguishable at short times, whereas discrepancies emerge at longer times, since the neglected quantum jumps in the non-Hermitian Hamiltonian case affect the convergence toward the steady state.
Moreover, the time scale over which the Lindblad and non-Hermitian Hamiltonian evolutions begin to significantly deviate is longer in the global approach, and the asymptotic trace distances are smaller.
These considerations imply that the no-jump conditional postselection has a stronger effect on the local master equation.
\begin{figure*}[ht] 
    \begin{center} \includegraphics[width=.7\linewidth]{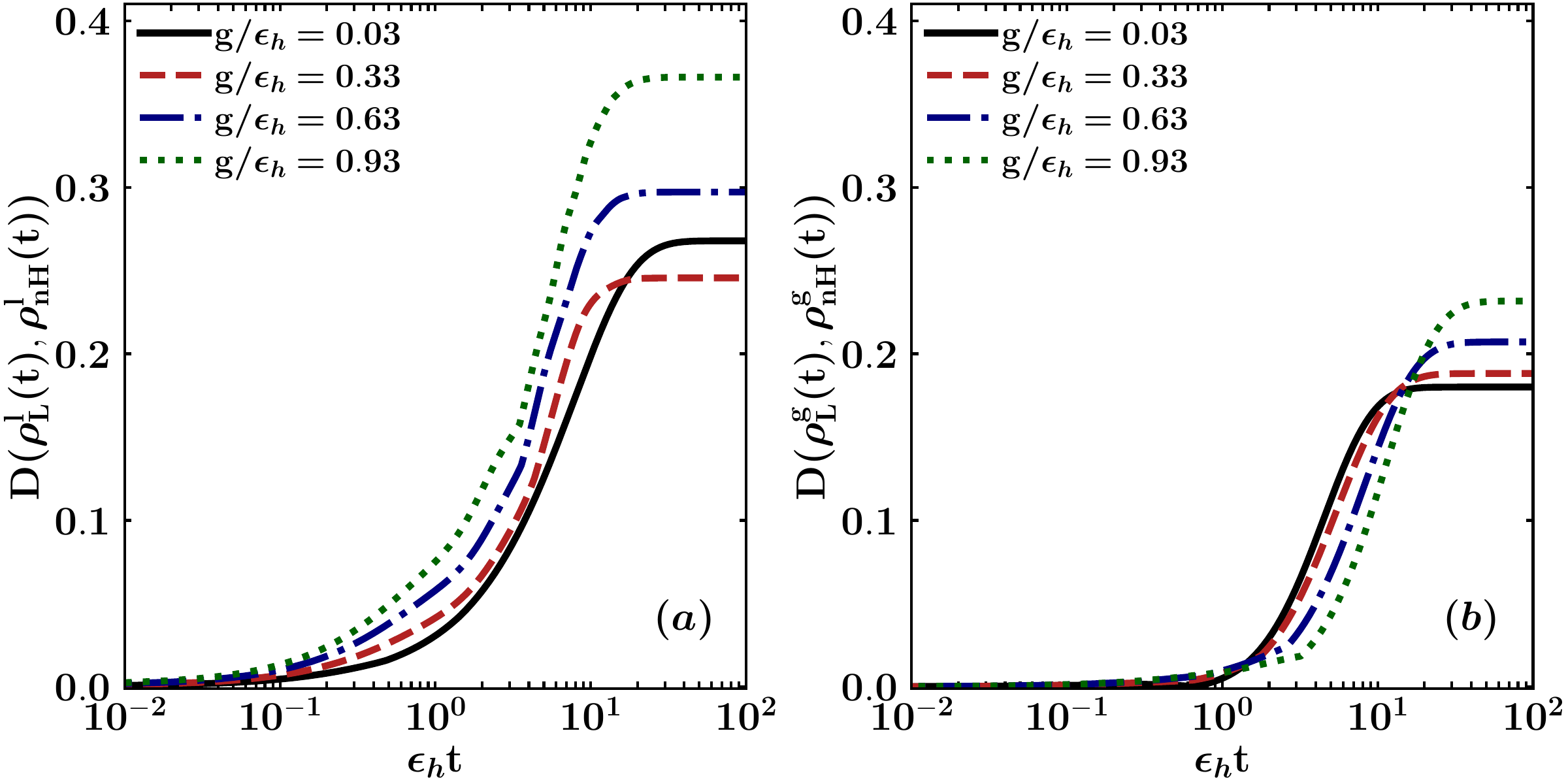} \caption{\label{fig:LnH}
    Comparison between the dynamics generated by Lindblad master equations and non-Hermitian Hamiltonians. We plot the
    trace distance in Eq.\,\eqref{eq:trdist} as a function of dimensionless time for four values of the coupling $g$. The Lindblad evolved state $\rho_L(t)$ is compared with the normalized non-Hermitian Hamiltonian state $\rho_{\rm nH}(t)$ for the local approach, as described by Eqs.\,\eqref{eq:dissipl} and \eqref{eq:nhl} (panel ($a$)), and for the global approach, as described by Eqs.\,\eqref{eq:dissipg} and \eqref{eq:nhg} (panel ($b$)).
    The parameters used in our simulations are:
    $\epsilon_c = \epsilon_h$, $\alpha_c = 0.2\epsilon_h$, $\alpha_h = 0.05\epsilon_h$, $T_c = 0.1\epsilon_h$, $T_h = \epsilon_h$, $\omega_c = 10\epsilon_h$. We tune the coupling parameter $g$ over the range $g \in [0.03,0.93] \epsilon_h$ (see legend). 
        }
    \end{center} 
\end{figure*}
}

\subsubsection{Thermodynamic considerations}
\label{sec:thermo-results}
The setting depicted in Fig.\,\ref{fig:model} describes a thermodynamic nonequilibrium configuration. In the Lindblad framework, the presence of a temperature bias gives rise to a steady state characterized by a constant heat current and a constant entropy production rate.
While this scenario has been extensively studied in the literature, its non-Hermitian Hamiltonian counterpart has not.

\begin{figure*}[ht]
    \centering
    \includegraphics[width=0.35\linewidth]{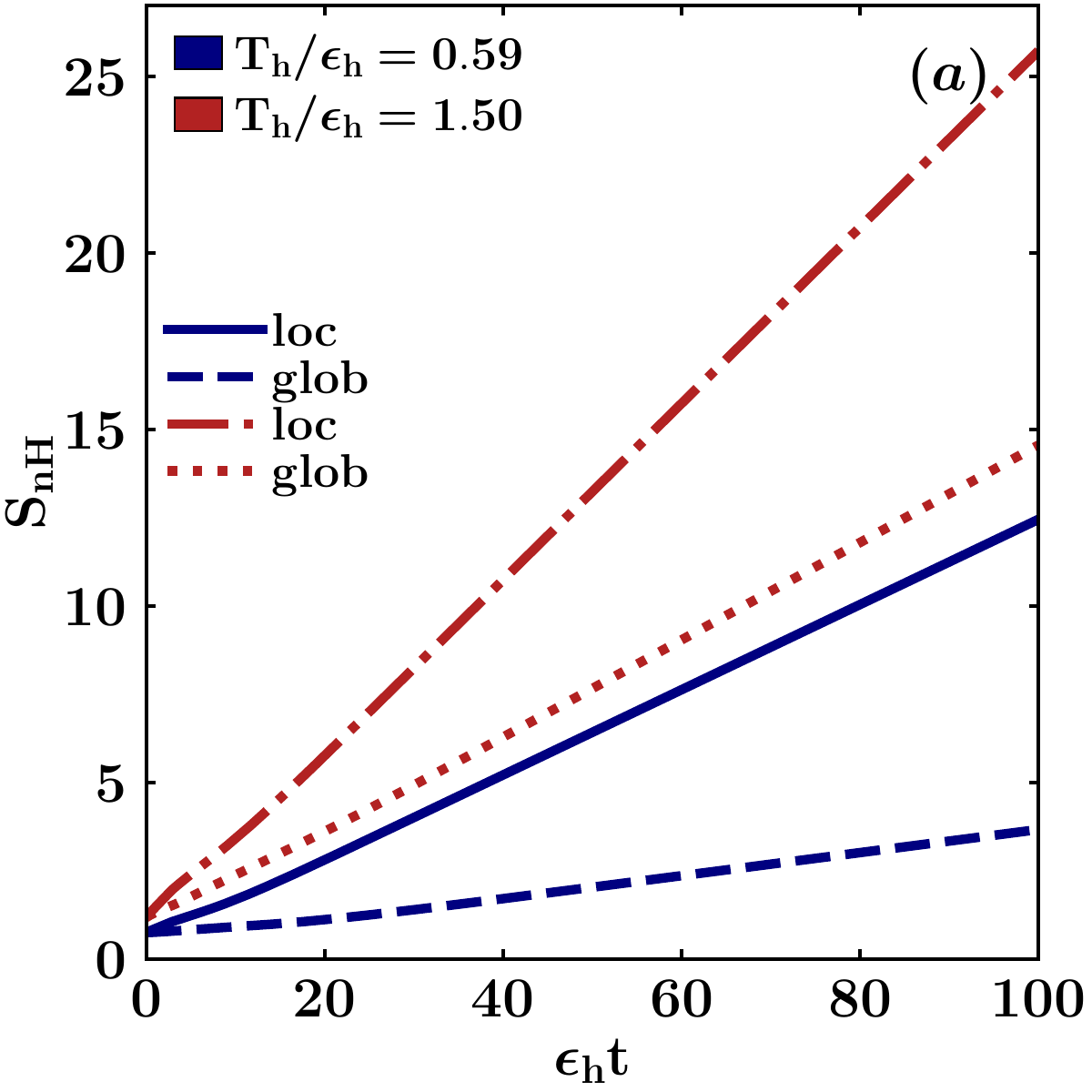}
    \includegraphics[width=0.35\linewidth]{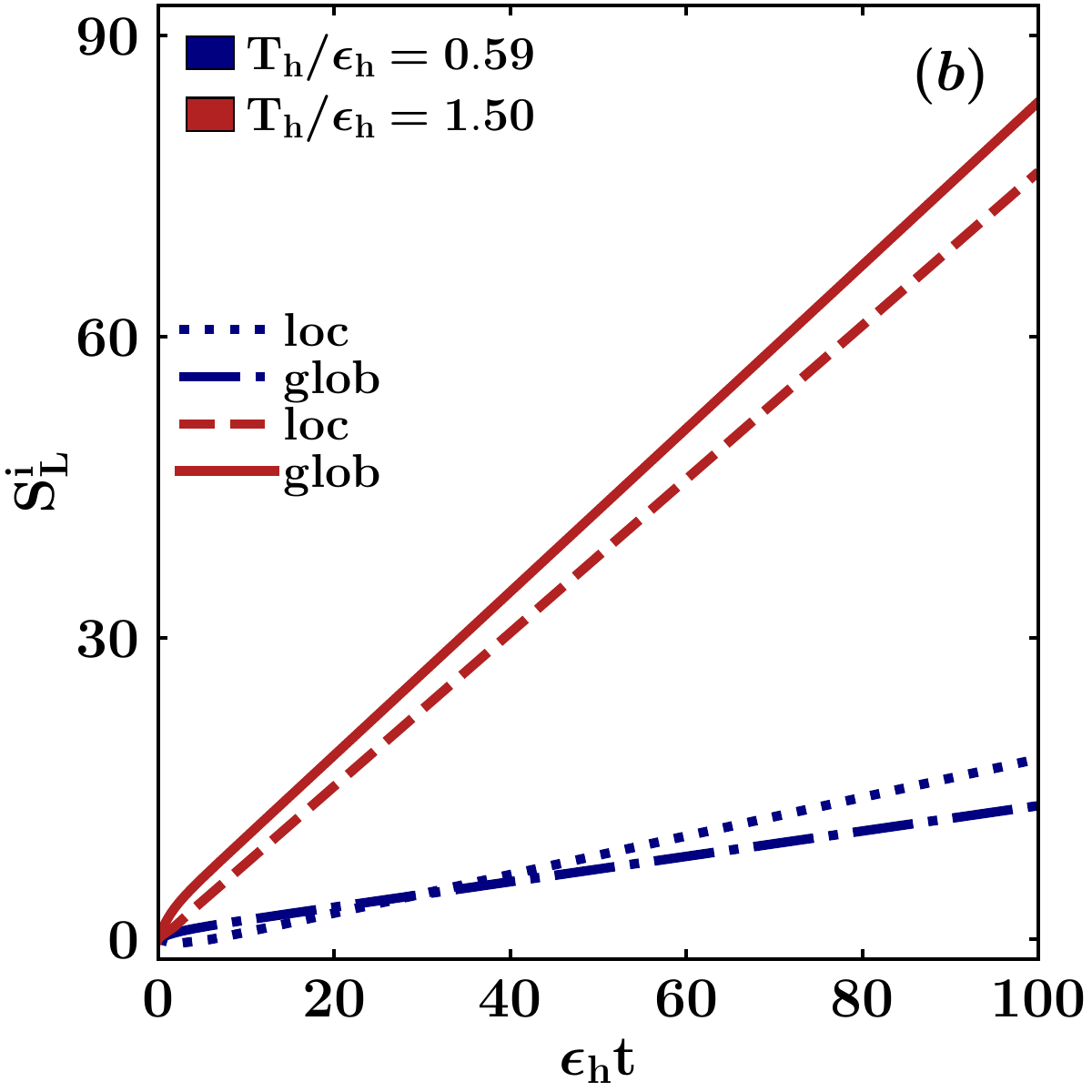}
    \caption{($a$) Non-Hermitian Hamiltonian von Neumann entropy and ($b$) irreversible entropy production in Lindblad dynamics as functions of dimensionless time for two values of the hot bath temperature $T_h$. 
    Results are shown for both local (loc) and global (glob) approaches. The parameters used in our simulations are: $\epsilon_c = \epsilon_h$, $\alpha_c = 0.2\epsilon_h$, $\alpha_h = 0.05\epsilon_h$, $T_c = 0.1\epsilon_h$, $\omega_c = 10\epsilon_h$, $g = 0.8\epsilon_h$. We tune the temperature of the hot bath $T_h \in \{0.59,1.50\}\epsilon_h$ (see legend). 
    }
    \label{fig:nHLentr}
\end{figure*}

Starting from the concepts introduced in Sec.\,\ref{sec:thermo-prel}, we notice that the second term in the non-Hermitian entropy production rate in Eq.\,\eqref{eq:nHentrprod2} 
captures deviations from the von Neumann entropy derivative and
incorporates temperature dependence when applied to our instance.
In the following numerical results, we fix the same initial thermal state with temperature $T_h$ as before. 
In Fig.\,\ref{fig:nHLentr}($a$) we plot the entropy $S_{\rm nH}$ in Eq.\,\eqref{eq:nHentr} as a function of time, for both local and global non-Hermitian Hamiltonian dynamics and for two different values of the hot bath temperature $T_h$. 
A qualitative correspondence between the non-Hermitian Hamiltonian entropy and the entropy production in Lindblad dynamics is apparent in Fig.\,\ref{fig:nHLentr}. For both local and global dynamics, $S_{\rm nH}$ in Fig.\,\ref{fig:nHLentr}($a$) and $S_{\rm L}^i$ in Fig.\,\ref{fig:nHLentr}($b$) display an approximately linear growth in time as one can expect for irreversible processes.


\subsection{
Exceptional points}\label{subsec:NHH_EP}
\subsubsection{Non-Hermitian Hamiltonian}
\begin{figure*} [ht]
    \begin{center} \includegraphics[width=.6\linewidth]
    {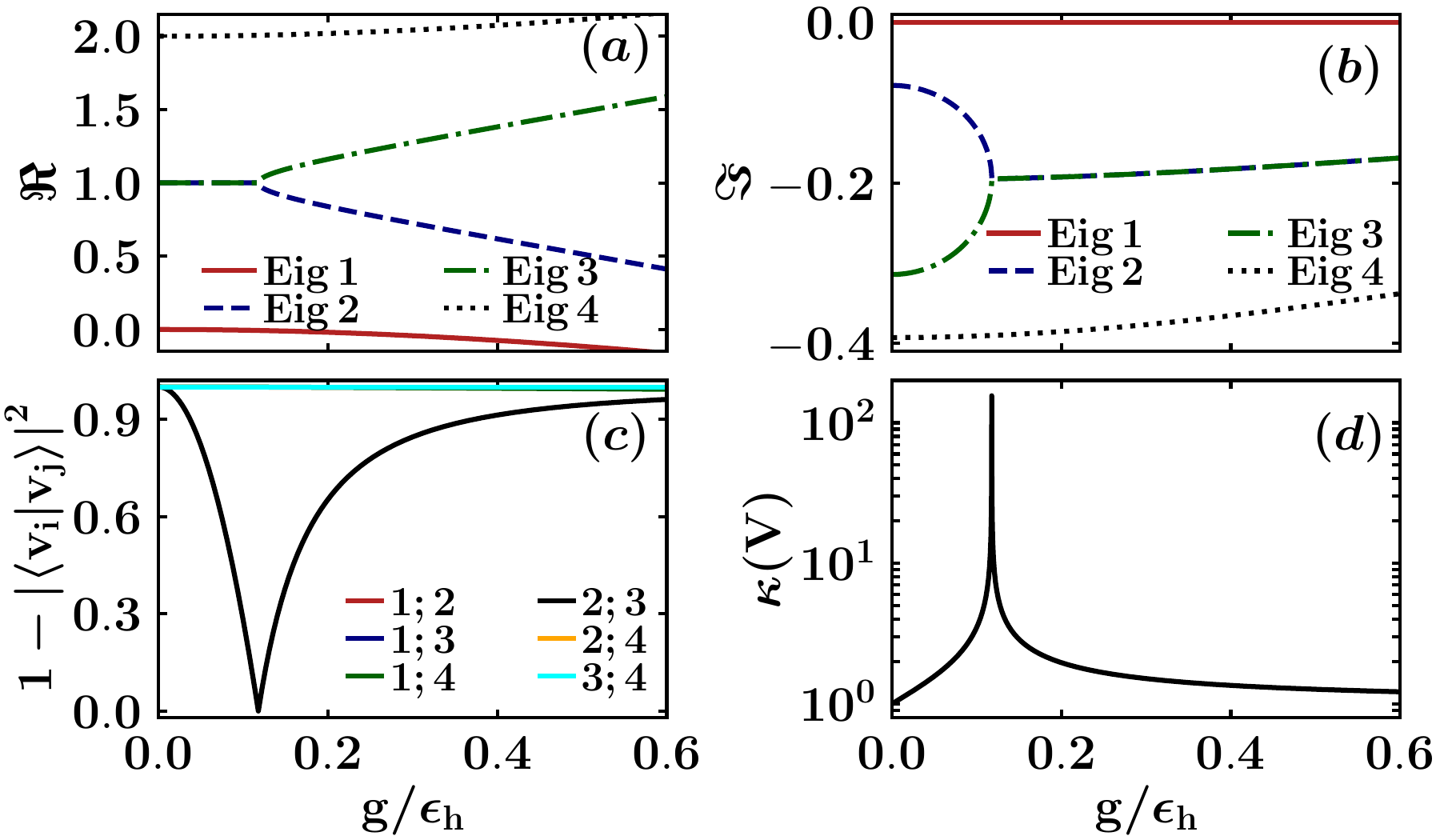} \caption{\label{fig:excpoints}
    Exceptional point in the non-Hermitian Hamiltonian derived from the local master equation.
    We report the coalescence of eigenvalues and eigenvectors 2 and 3. This is shown by examining the real part in panel ($a$) and the imaginary part in panel ($b$) of the eigenvalues as functions of $g$ (in units of $\epsilon_h$). Non-orthogonality of eigenvectors 2 and 3 is computed by evaluating $1 - |\langle v_i | v_j \rangle|^2$ for each pair of right eigenvectors $i,j$ as a function of $g$ (in units of $\epsilon_h$) and shown in panel ($c$). Divergence of the condition number of the eigenvector matrix $\kappa(V)$ is shown in panel ($d$) as a function of the coupling $g$ (in units of $\epsilon_h$). The parameters used in our simulations are:
    $\epsilon_c = \epsilon_h$, $\alpha_c = 0.2\epsilon_h$, $\alpha_h = 0.05\epsilon_h$, $T_c = 0.1\epsilon_h$, $T_h = \epsilon_h$, $\omega_c = 10\epsilon_h$. We tune the coupling $ g $ over the range $g \in [0.0,0.6] \epsilon_h $ and find $\mathrm{EP}$ at $g \approx 0.12\epsilon_h$.}
    \end{center} 
\end{figure*}
We now investigate the occurrence of exceptional points, starting from the non-Hermitian Hamiltonian descriptions.
In our analysis, we observe the emergence of exceptional points when increasing the coupling strength $g$ between the two qubits, but only within the local approach and when the qubit energy detuning is zero ($\delta = 0$). 
This suggests that the relevant non-Hermitian Hamiltonian features of the system depend crucially on the structure of the jump operators. 
In this regard, it is useful to check the necessary condition for the existence of exceptional points 
\begin{equation}
\label{nonCommHGamma}
[\mathcal{H},\Gamma]\neq 0
\end{equation}
\cite{moiseyev2011non}, where $\mathcal{H}$ and $\Gamma$ are the Hermitian operators in the general form of a non-Hermitian Hamiltonian as in Eq.\,\eqref{eq:nheq}. 
In the case of the global approach, for our choice of internal interaction $g\sigma_x^h\sigma_x^c$ between the qubits, the resulting commutator vanishes for all parameter values. This means that the eigenvectors constitute an orthogonal set, which in turn rules out the appearance of exceptional points, consistent with what has been found in earlier studies across various settings \cite{khandelwal2021signatures,khandelwal2025emergent,svegborn2026framework}. 
\begin{widetext}
In contrast, within the local approach, the commutator has a structure in which only the antidiagonal elements are nonzero:
\begin{equation}
[\mathcal{H},\Gamma^l]=
\frac{1}{2} g
\begin{pmatrix}
0 & 0 & 0 & \gamma_c(\epsilon_c) + \gamma_h(\epsilon_h) \\
0 & 0 & -\gamma_c(\epsilon_c) + \gamma_h(\epsilon_h) & 0 \\
0 & \gamma_c(\epsilon_c) - \gamma_h(\epsilon_h) & 0 & 0 \\
-\gamma_c(\epsilon_c) - \gamma_h(\epsilon_h) & 0 & 0 & 0
\end{pmatrix}.
\end{equation}
\end{widetext}
The commutator vanishes in case of $g=0$, i.e., when the two qubits are decoupled.
For $g=0$, each qubit relaxes to the Gibbs state of the bath it is coupled to. This indicates that, in this setup, nonequilibrium conditions are essential for the emergence of exceptional points. 
However, since Eq.\,\eqref{nonCommHGamma} is only a necessary condition, there can exist parameter sets for which the commutator is nonzero but no exceptional points appear. In such cases, the eigenvalues may become degenerate while the corresponding eigenvectors remain linearly independent. This occurs, for instance, in the local approach when $\epsilon_c\neq \epsilon_h$. 
Notably, for selected nonzero values of $g$, exceptional points do emerge. 
Considering the local non-Hermitian Hamiltonian, we report eigenvalue coalescence in Fig.\,\ref{fig:excpoints}($a$) and ($b$).
The complex eigenvalues merge at specific values of $g$ that depend on the coupling strengths to the baths $\alpha_j$. 
This can be tracked by analyzing the behavior of their real parts [Fig.\,\ref{fig:excpoints}($a$)] and imaginary parts [Fig.\,\ref{fig:excpoints}($b$)] as $g$ increases.
Corresponding eigenvector coalescences and divergence of the condition number are displayed in
Fig.\,\ref{fig:excpoints}($c$) and ($d$), respectively. See Appendix \ref{app:ep_local_validity} for the corresponding result in a parameter regime where the local approach is microscopically valid \cite{cattaneo2019local}.

\begin{figure*}[ht]
    \centering
    \includegraphics[scale=0.3]{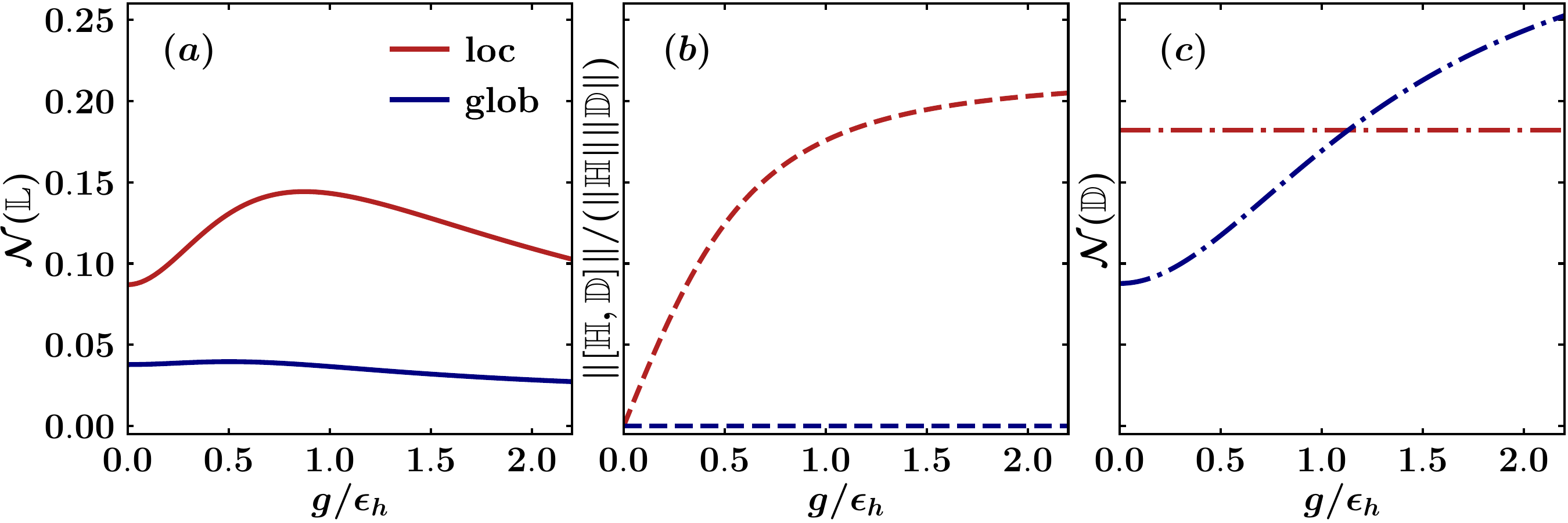}
    \caption{
    (a)
    Non-normality of the local (loc) and global (glob) Lindbladians, defined in Eq.\,\eqref{eq:nonnormality}, shown as a function of the coupling $g$ (in units of $\epsilon_h$). (b) The $[\mathbb{H},\mathbb{D}]$ contribution, $\left\Vert [\mathbb{H},\mathbb{D}] \right\Vert / (\left\Vert \mathbb{H} \right\Vert \,\left\Vert \mathbb{D} \right\Vert)$, as a function of $g/\epsilon_h$. (c) Non-normality of the dissipative part of the Lindbladian, $\mathbb{D}$, as a function of $g/\epsilon_h$. The parameters used in our simulations are: $\epsilon_c = \epsilon_h$, $\alpha_c = 0.2\epsilon_h$, $\alpha_h = 0.05\epsilon_h$, $T_c = 0.1\epsilon_h$, $T_h=\epsilon_h$, $\omega_c = 10\epsilon_h$. We tune the coupling $g$ over the range $[0.0,2.2] \epsilon_h$.}
    \label{fig:nonnorm}
\end{figure*}

\subsubsection{Lindblad and hybrid postselection frameworks}
We now include the quantum-jump terms, thereby forming Lindblad dissipators, and study the emergence of Liouvillian exceptional points.
While local and global Lindblad master equations have been extensively studied in the literature, their potential for generating exceptional points remains largely unexplored.
Here, we investigate this aspect for the same parameter regime considered in the previous section.
Analogously to the non-Hermitian Hamiltonian case, we observe exceptional points only for the local master equation.

Adapting previous reasoning, a {Lindblad master equation} when vectorized (Liouville space representation) is
$
|\dot{\rho}) = \mathbb{L}|\rho),
$
where $\mathbb{L}$ is a (typically non-Hermitian) {Liouvillian matrix} acting on a Hilbert space of dimension $d^2$, $d$ being the system Hilbert space dimension.
An exceptional point appears when two (or more) eigenvectors coalesce, i.e. the operator becomes 
defective (non-diagonalizable).
This can only occur when $\mathbb{L}$ is \textit{non-normal}, that is:
\begin{equation}
\label{NOcommL}
[\mathbb{L}, \mathbb{L}^\dagger] \neq 0\,.
\end{equation}

If $\mathbb{L}$ is \textit{normal}, i.e.
$
[\mathbb{L}, \mathbb{L}^\dagger] = 0,
$
its eigenvectors form an orthogonal set.
Therefore, Eq.\,\eqref{NOcommL} is a necessary condition for the emergence of exceptional points.
Let us split the Liouvillian as
\begin{equation}\label{eq:Lindbladian_L}
\mathbb{L} = \mathbb{H} + \mathbb{D}\,,
\end{equation}
where $\mathbb{H}$ is the matrix representation of $-i[H,\,\cdot\,]$ (skew-Hermitian, unitary part) 
and
$\mathbb{D}$ is the dissipative part. 
Accordingly, the commutator in Eq.\,\eqref{NOcommL} can be written as
\begin{equation}
[\mathbb{L}, \mathbb{L}^\dagger] = [\mathbb{H} + \mathbb{D},\, -\mathbb{H} + \mathbb{D}^\dagger]
    = [\mathbb{D}, \mathbb{D}^\dagger] + [\mathbb{H},\, \mathbb{D} + \mathbb{D}^\dagger].    
\end{equation}
Hence, a sufficient (but not necessary) condition for normality is
\begin{equation}
\label{CommHD}
[\mathbb{H}, \mathbb{D}] = 0
\qquad \text{and} \qquad
[\mathbb{D}, \mathbb{D}^\dagger] = 0\,. 
\end{equation}
If these commutators vanish, $\mathbb{L}$ is normal implying no exceptional points.
Motivated by these observations, we quantify the \enquote{non-normality} of the Lindbladian via,
\begin{equation}\label{eq:nonnormality}
\mathcal{N}(\mathbb{L}) = \frac{\left\Vert [\mathbb{L}, \mathbb{L}^\dagger] \right\Vert}{\left\Vert \mathbb{L} \right\Vert^{2}}\,,
\end{equation}
using for simplicity the Frobenius norm.
Fig.\,\ref{fig:nonnorm}($a$) shows the non-normality $\mathcal{N}(\mathbb{L})$, defined in Eq.\,\eqref{eq:nonnormality}, for the Lindbladian associated with the local and global approaches. We observe that the local approach exhibits a larger non-normality than the global approach for the same choice of parameters as in Fig.\,\ref{fig:excpoints}, over the entire range of coupling strengths $g$.
In Fig.\,\ref{fig:nonnorm}($b$) we show the contribution to the Lindbladian non-normality arising from the commutator $[\mathbb{H},\mathbb{D}]$, which is analogous to the condition $[\mathcal{H},\Gamma]\neq 0$ that was necessary for exceptional points in the non-Hermitian Hamiltonian dynamics discussed in Sec.\,\ref{subsec:NHH_EP}. In the global approach this contribution vanishes.
In this regard, we remind that in general within a {global Lindblad master equation} the dissipators are constructed from the {eigenoperators $A_\omega$ (the jump operators) of the system Hamiltonian}, namely
$
[\mathcal{H}, A_{\omega}] = -\omega A_{\omega}
$
\cite{breuer2007theory}.
On the contrary, in the local approach the commutator $[\mathbb{H},\mathbb{D}]$ is nonzero for all $g\neq 0$ and increases monotonically across the explored parameter range.
Fig.\,\ref{fig:nonnorm}(c) shows instead the non-normality of the dissipative part $\mathbb{D}$ of the Lindbladian in Eq.\,\eqref{eq:Lindbladian_L}, defined analogously as
\begin{equation}
\mathcal{N}(\mathbb{D}) = \frac{\left\Vert [\mathbb{D}, \mathbb{D}^\dagger] \right\Vert}{\left\Vert \mathbb{D} \right\Vert^{2}}\,.
\end{equation}
For this quantity, the local result is independent of $g$, while the global result is always nonzero and increases with $g$, with a change in concavity. Therefore, within the global approach the only contribution that makes the Lindbladian non-normal comes from the non-normality of the dissipator (i.e., the term including quantum jumps).
\begin{figure*} [ht!]
    \centering
    \includegraphics[width=0.24\linewidth]{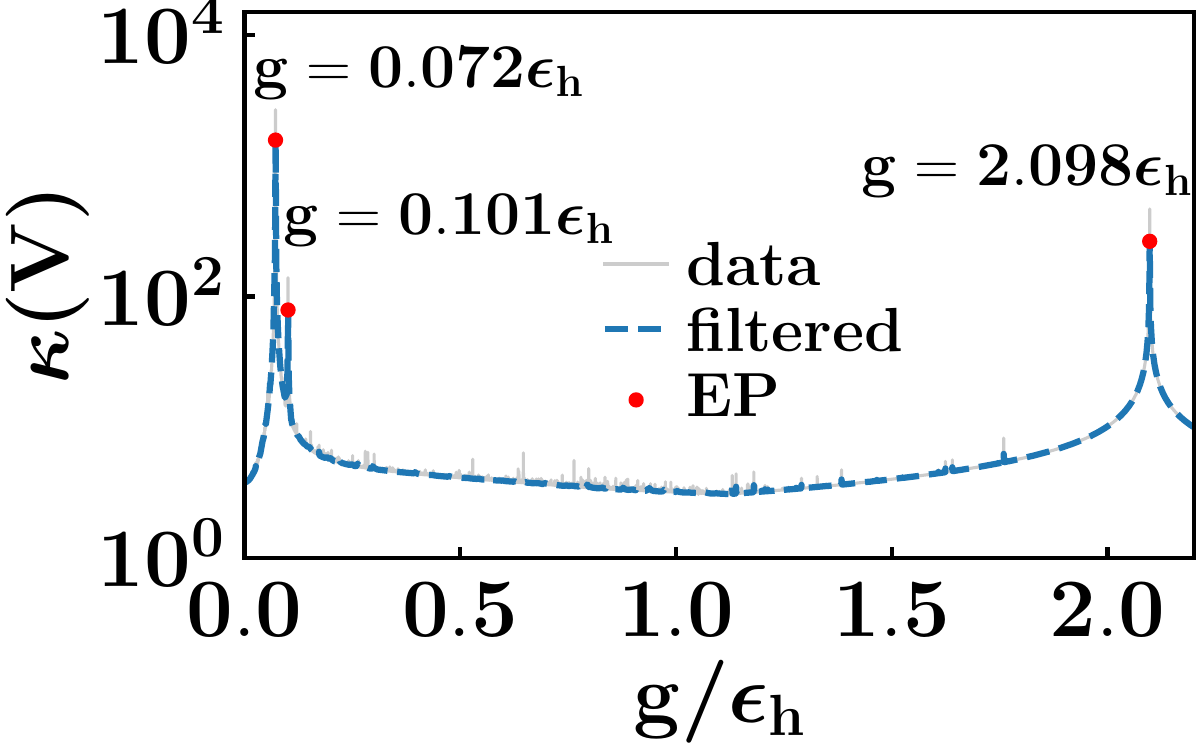}%
    \includegraphics[width=0.24\linewidth]{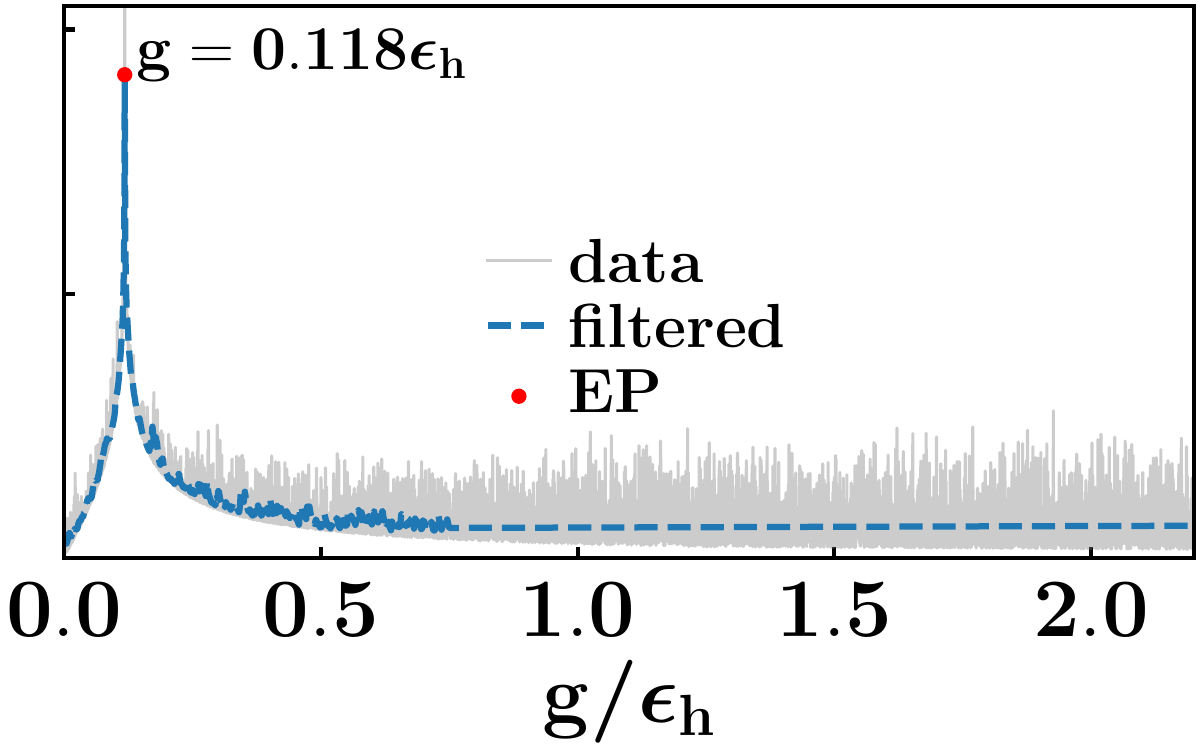}%
    \includegraphics[width=0.24\linewidth]{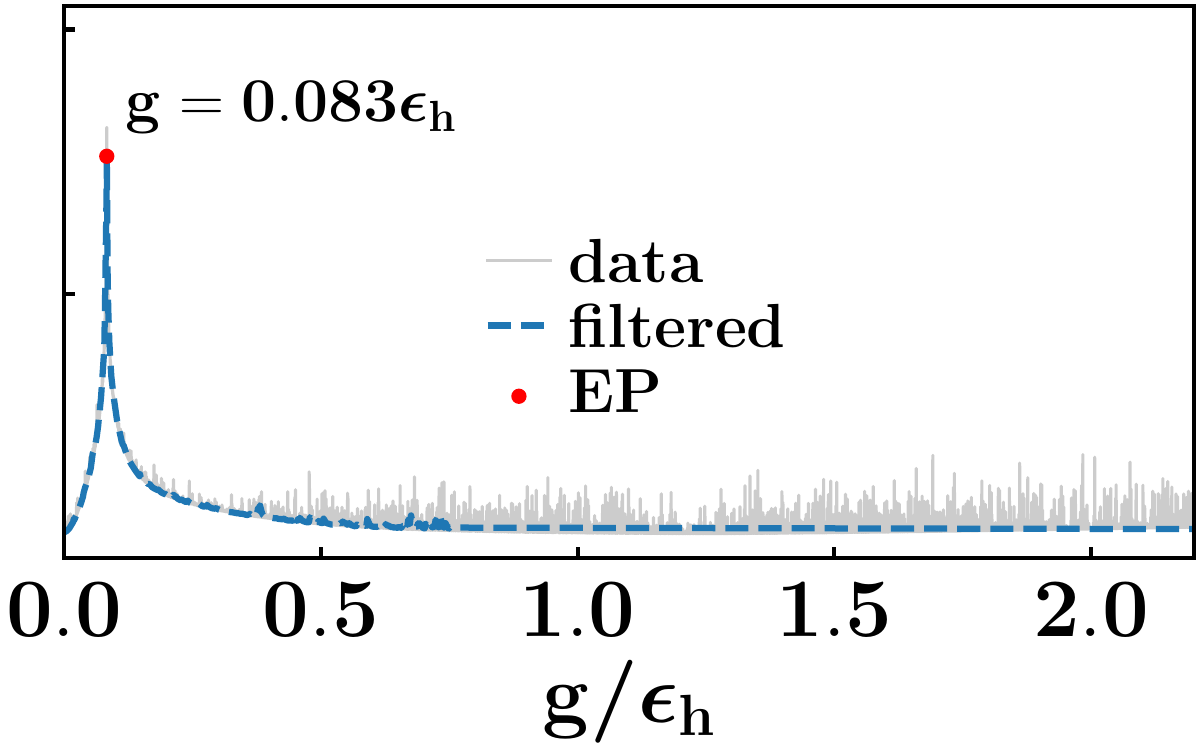}%
    \includegraphics[width=0.24\linewidth]{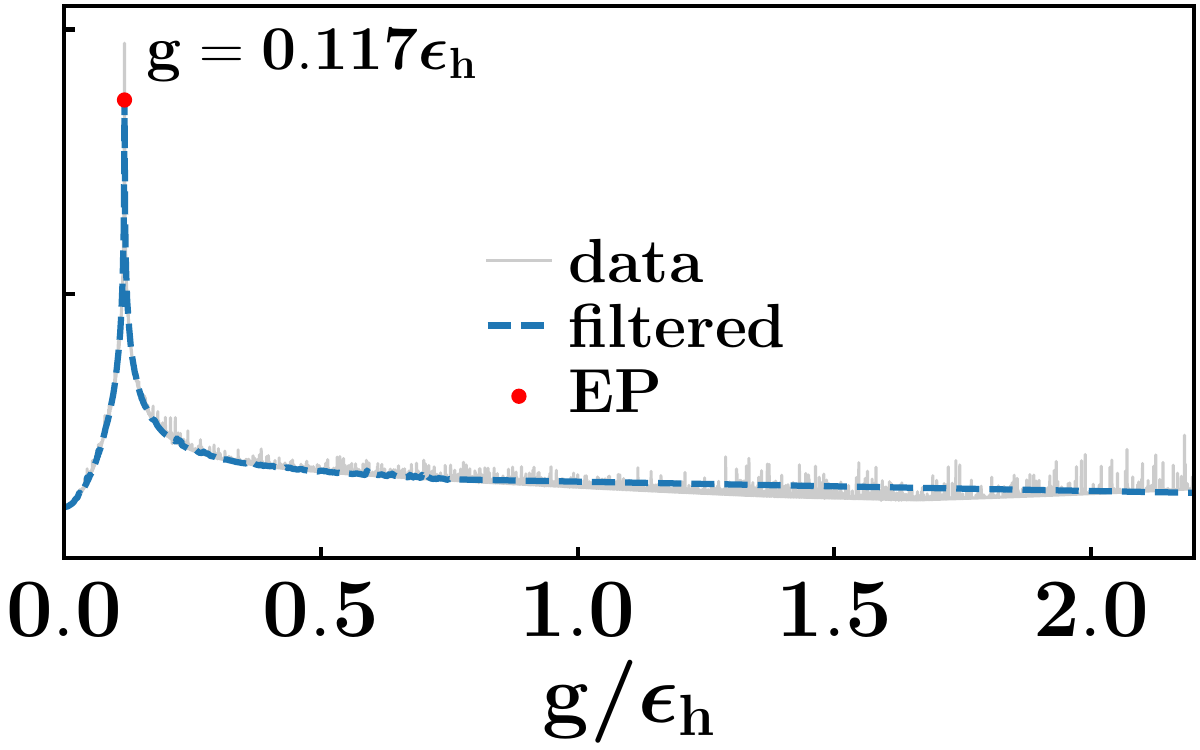}\\[4pt] 

    \includegraphics[width=0.24\linewidth]{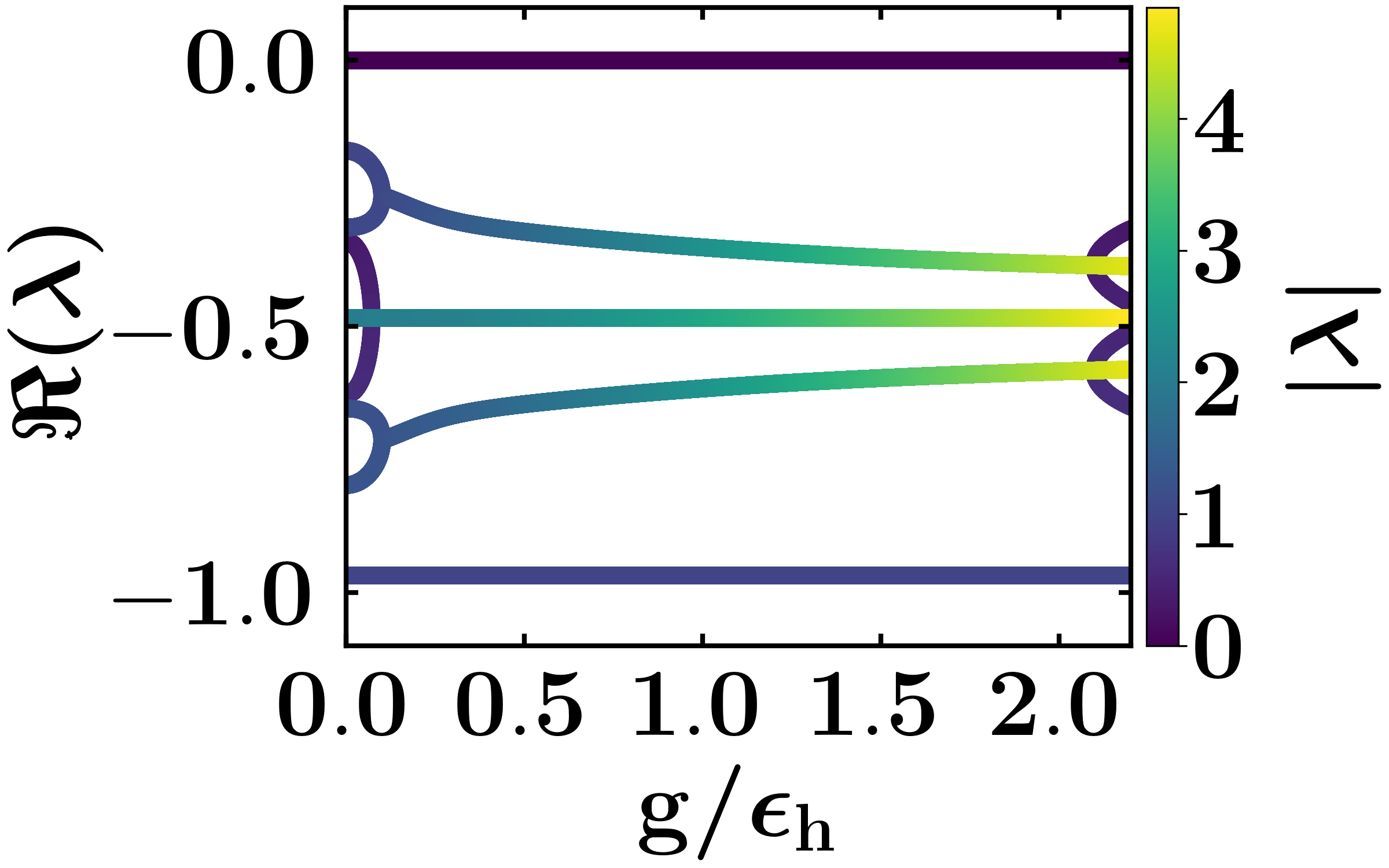}%
    \includegraphics[width=0.24\linewidth]{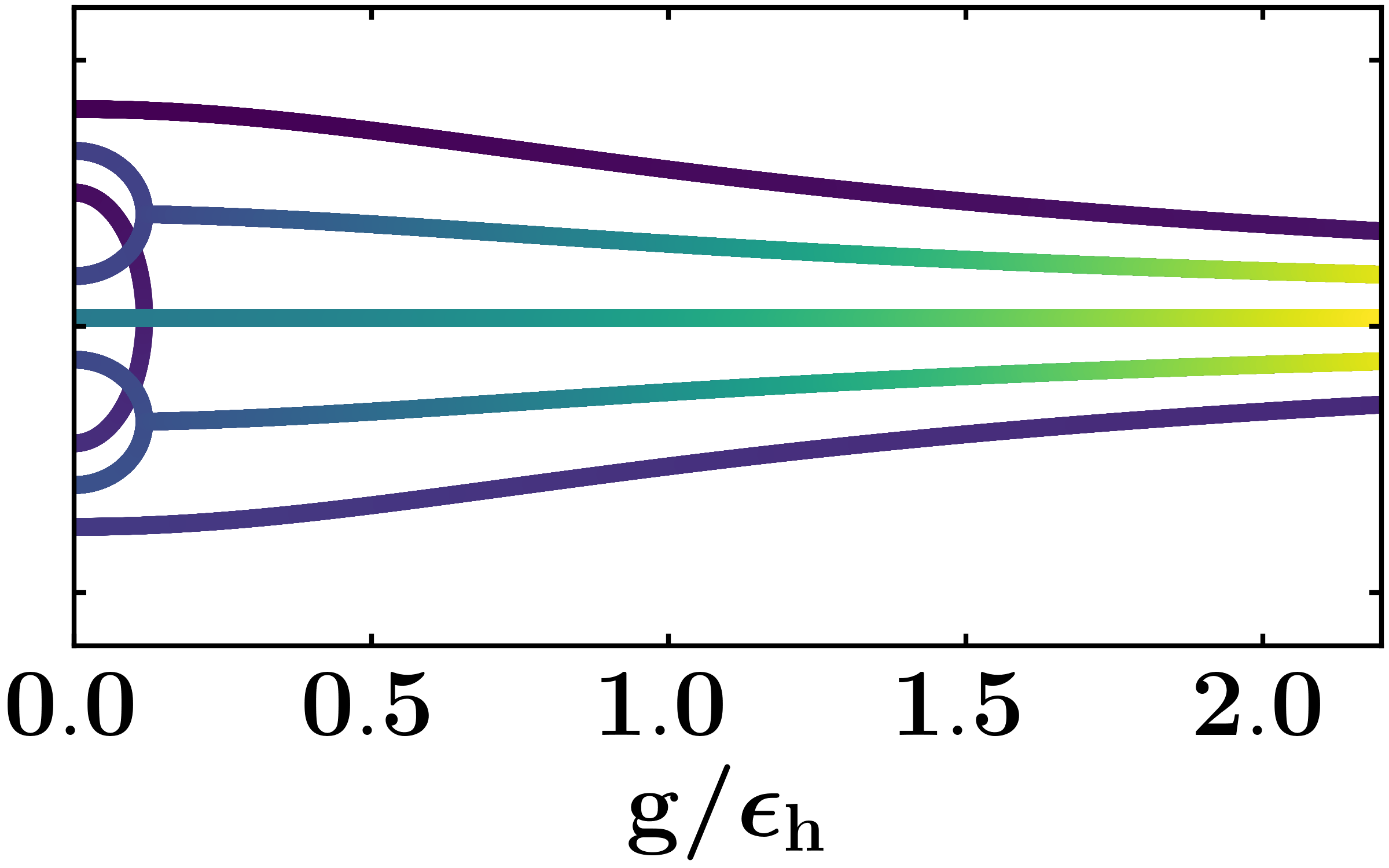}%
    \includegraphics[width=0.24\linewidth]{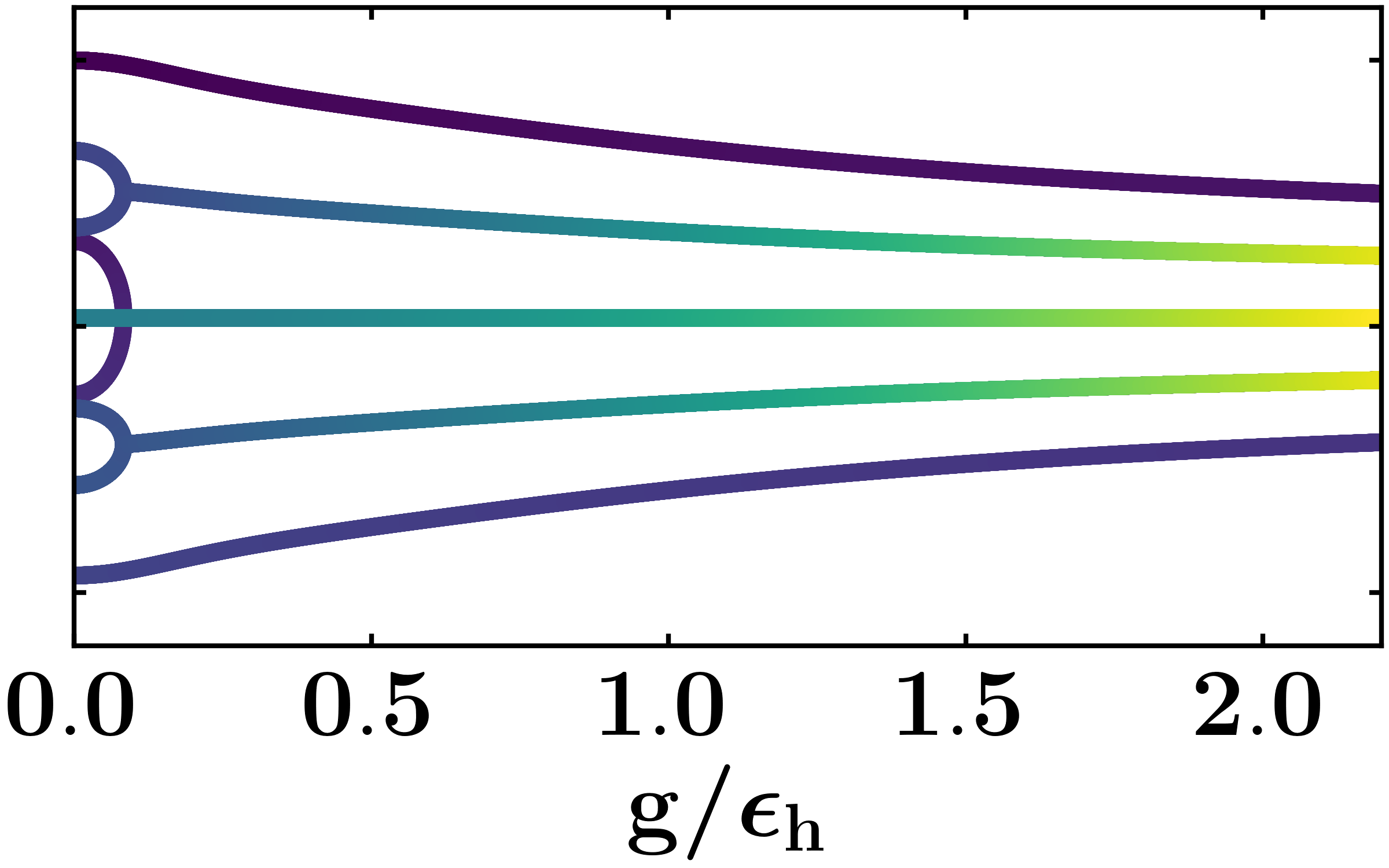}%
    \includegraphics[width=0.24\linewidth]{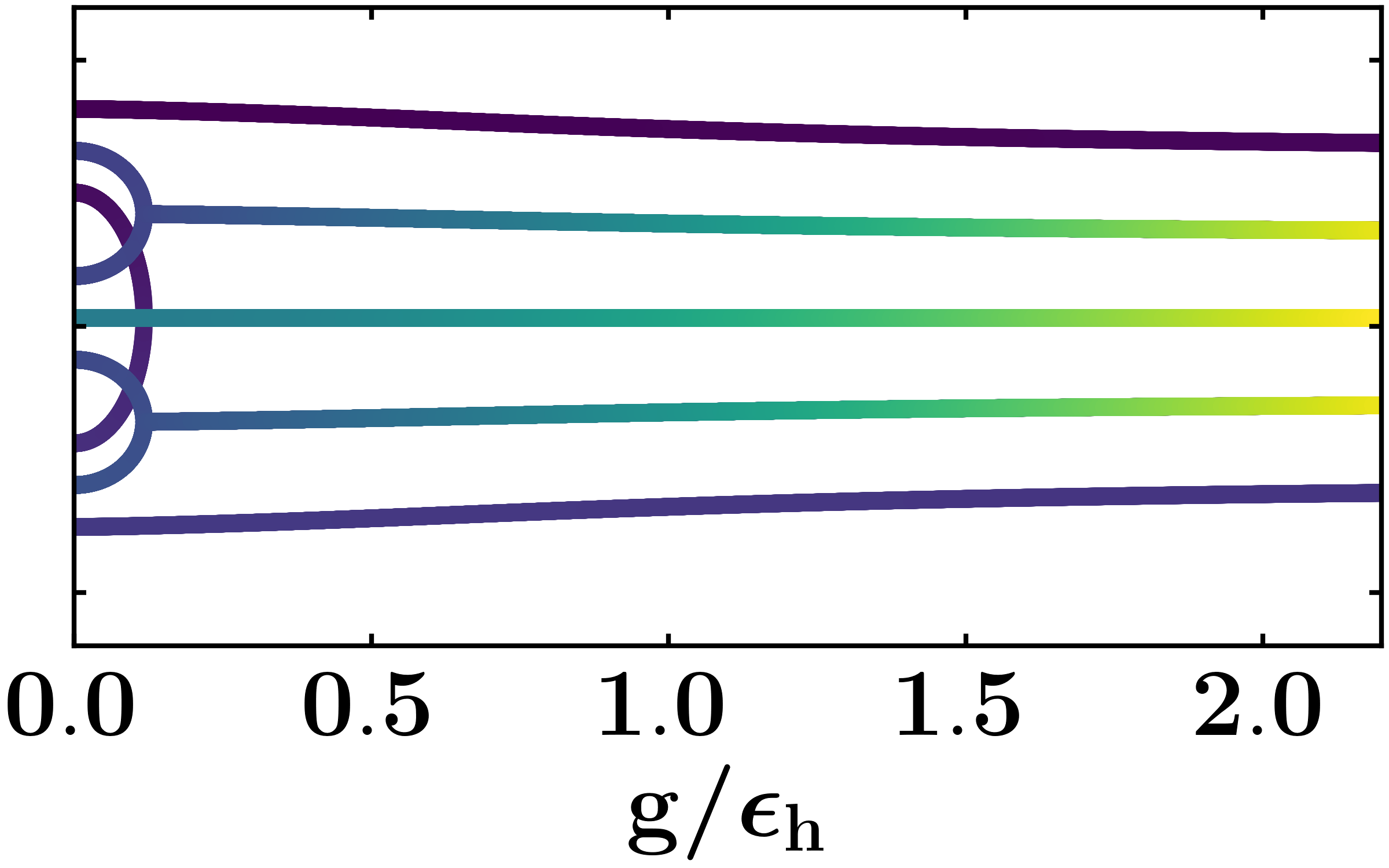} 
    \caption{
    Liouvillian exceptional points derived from the local master equation.
    Upper row: divergence of the condition number of the right eigenvectors $\kappa(V)$ as a function of the coupling $g$ (in units of $\epsilon_h$). The data have been filtered to remove spurious peaks caused by numerical instability.
    Lower row: real parts of the eigenvalues as functions of $g$ (in units of $\epsilon_h$), showing the coalescence of eigenvalues. The color bar indicates the modulus $|\lambda|$ of the corresponding right eigenvalues. Columns correspond to: (1) local Lindblad dynamics, (2) local non-Hermitian Hamiltonian dynamics (Liouvillian without quantum jumps), (3) hybrid case with postselection on the cold bath in the local approach, and (4) hybrid case with postselection on the hot bath in the local approach.
    The parameters used in our simulations are:
    $\epsilon_c = \epsilon_h$, $\alpha_c = 0.2\epsilon_h$, $\alpha_h = 0.05\epsilon_h$, $T_c = 0.1\epsilon_h$, $T_h = \epsilon_h$, $\omega_c = 10\epsilon_h$. We tune the coupling $ g $ over the range $g \in [0.0,2.2] \epsilon_h $.}
    \label{fig:condnumber_repart}
\end{figure*}
In Appendix \ref{app:nonnormality_max} we report the same quantities for a different set of parameters, where the behavior of the local and global approaches is reversed for the non-normality of the full Lindbladian. Even in that case, $\mathcal{N}(\mathbb{D})$ remains constant for the local approach and increases for the global one, while the $[\mathbb{H},\mathbb{D}]$ contribution again vanishes in the global approach and increases with $g$ in the local approach.

The fact that the local approach is always non-normal is consistent with what we are about to observe: Liouvillian exceptional points are typically present. By contrast, the smaller non-normality in the global approach does not by itself guarantee the absence of exceptional points; nevertheless, we do not observe any in this case.

This highlights a potentially crucial role of the commutator $[\mathbb{H},\mathbb{D}]$ in determining the presence or absence of exceptional points.
To see this explicitly, we numerically diagonalize the Liouvillian corresponding to the Lindbladian to obtain its eigenvalues and eigenvectors, and we compute the condition number of the eigenvector matrix.
For completeness, we do the same for the Liouvillians corresponding to the non-Hermitian Hamiltonian dynamics (matrix representations of $\mathcal{L}'_{l}$ and $\mathcal{L}'_{g}$).
We also perform analogous analyses for selected hybrid configurations combining Lindblad and non-Hermitian Hamiltonian descriptions.
For brevity, we present numerical results only for the local approach, since the global one does not exhibit any exceptional points in our analysis (data not shown).
We show in Fig.\,\ref{fig:condnumber_repart} the divergence of the condition number (see upper row of panels) and the crossing of the real parts of the eigenvalues (lower row) for four different cases in the local approach. The first column refers to the Lindblad dynamics, where three exceptional points appear, with the central one being very close to the corresponding point in the non-Hermitian Hamiltonian case. The second column again shows the non-Hermitian Hamiltonian dynamics, but here we diagonalize the Liouvillian without including quantum jumps. The peak occurs at the same coupling $g$ as in Fig.\,\ref{fig:excpoints}, but with more crossings of the real parts of the eigenvalues, as the Liouvillian now has 16 eigenvalues. 
The third column corresponds to the case where we postselect only on the cold bath.
Hence, in this hybrid setting the effective interaction with the cold bath is described by a non-Hermitian Hamiltonian term and the one with the hot bath by a Lindblad dissipator.
The fourth column refers instead to postselection only on the hot bath, reversing the previous setting.
In both hybrid cases, we observe a single peak at a $g$ value close to that of the non-Hermitian Hamiltonian case. Moreover, the spectrum of these hybrid situations shares features with both the non-Hermitian Hamiltonian and the Lindblad spectra.

\section{Conclusions}\label{sec:conclusions}
In summary, we studied the local and global Markovian master equations under the perspective of non-Hermitian physics.
This work is motivated by the observation that local and global master-equation approaches are well established within the Lindblad formalism. In contrast, their non-Hermitian Hamiltonian counterparts remain largely unexplored, together with characteristic features of non-Hermitian physics such as exceptional points, even in the Lindblad setting.
We considered the paradigmatic nonequilibrium configuration consisting of a two-qubit system coupled to two thermal baths at different temperatures.
A key feature of the model is its minimality, which facilitates the understanding of the essential concepts.

We first observed that
while non-Hermitian Hamiltonian and Lindblad modelings yield similar short-time dynamics, their long-time behaviors diverge, leading to different steady states. These differences become more pronounced with increasing inter-qubit coupling.
From a thermodynamic perspective, we analyzed the dynamics of non-Hermitian entropies, which successfully retain sensitivity to the temperature bias, thus detecting irreversible dynamics.
Notably, we showed that the entropy production rate reaches different steady values, signaling a nonequilibrium steady state that differs between the two approaches.

As a key result, we reported the emergence of exceptional points in the spectrum of the effective non-Hermitian Hamiltonians.
Exceptional points are possible only for the dynamics derived from the 
local master equation for selected nonzero values of the inter-qubit coupling that sets the degree of nonequilibrium.
We extended our analysis on exceptional points to the Lindblad case, including hybrid settings where the interaction with one bath is modeled with a non-Hermitian Hamiltonian term and the one with the other bath via a Lindblad dissipator.
Analogously to the non-Hermitian Hamiltonian case, exceptional points emerge only for the local master equation. We showed that the number of exceptional points increases when considering the full Lindbladian, if compared to the non-Hermitian Hamiltonian setting and hybrid configurations.

Overall, our results contribute understanding non-Hermitian physics in nonequilibrium thermodynamic settings across different master-equation descriptions.
They may also be amenable to experimental realization in circuit-QED platforms \cite{naghiloo2019quantum, ronzani2018tunable}.
Furthermore, several theoretical aspects open up promising avenues for future research.
They include performing similar analyses for the Redfield master equation \cite{redfield1957theory, ishizaki2009adequacy}, widely used in chemical physics, and its regularized forms \cite{PhysRevA.100.012107,d2023time, PhysRevX.14.031010}.
Another exciting avenue is constituted by finding links with thermodynamic precision, for instance in the framework of thermodynamic uncertainty relations for local and global master equations \cite{PhysRevE.109.034112}. 
Additionally, it would be interesting to explore the subtle role of the Lamb shift term \cite{correa2025potential} in the generation of exceptional points, as well as regimes beyond master equations \cite{khandelwal2025emergent}. 

\section*{Acknowledgments}
D.F. thanks V. Cavina, S. Cusumano and F. E. Quintela Rodriguez for fruitful discussions.
G.D.B. acknowledges funding from IQARO (Spin-orbitronic Quantum Bits in Reconfigurable 2DOxides) project of the European Union’s Horizon Europe research and innovation programme under grant agreement n. 101115190. D.F. acknowledges financial support from PNRR MUR Project No. PE0000023-NQSTI and from University of Catania via PNRR-MUR Starting Grant project PE0000023-NQSTI.


\appendix
\section{Lindblad and non-Hermitian Hamiltonian dynamics approaching steady states}\label{app:trace_dist_ss}
\begin{figure}[h] 
    \centering
    \includegraphics[scale=0.34]{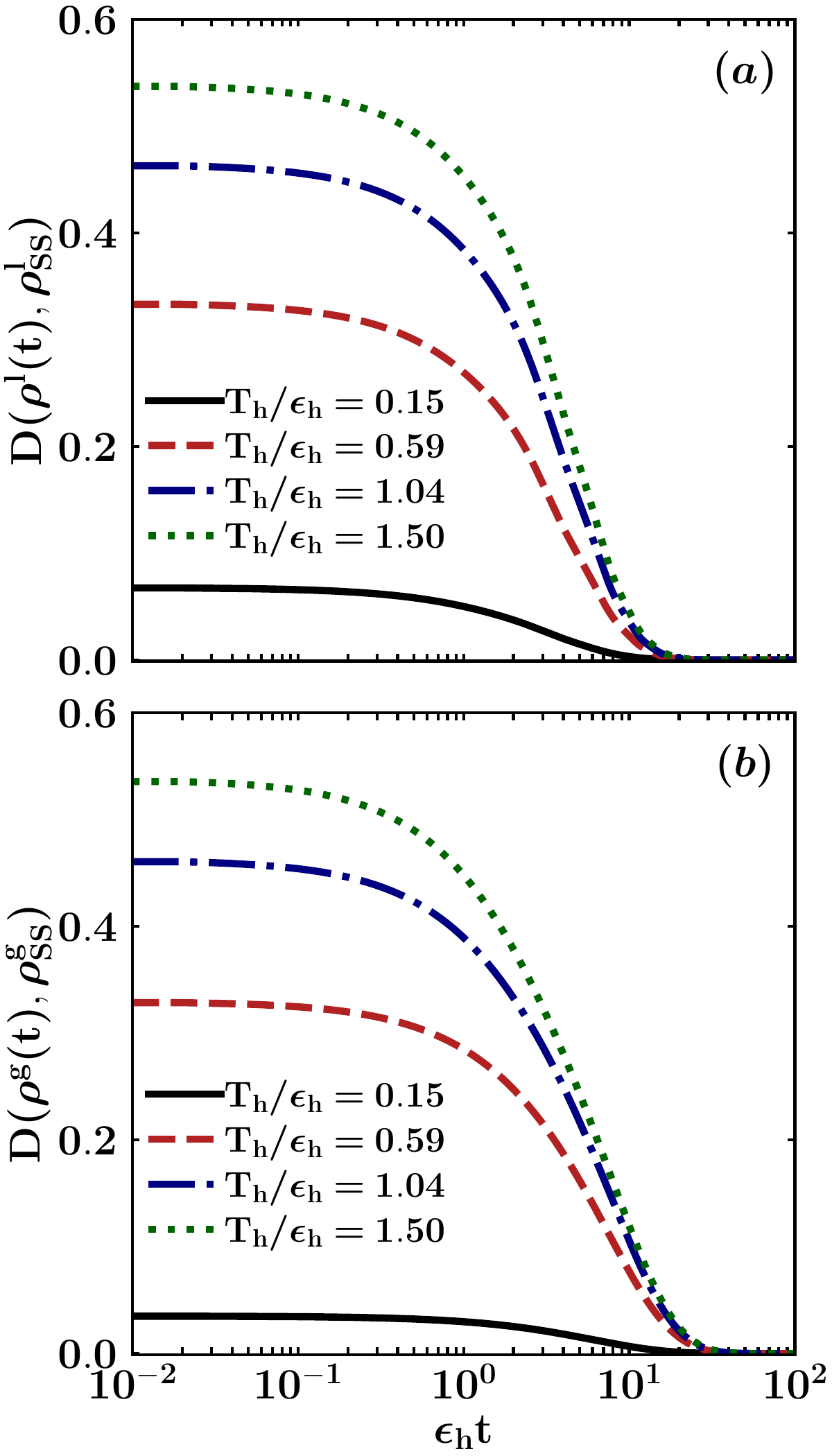} \caption{\label{fig:NH_lg_SS}
    Relaxation to the steady state in non-Hermitian Hamiltonian dynamics.
    Trace distance in Eq.\,\eqref{eq:trdist} as a function of dimensionless time for four values of the temperature of the hot bath $T_h$. The normalized non-Hermitian Hamiltonian evolved state $\rho_{\rm nH}(t)$ for the local approach (see Eq.\,\eqref{eq:nhl}) is compared with the longest-lived state in panel ($a$). The same non-Hermitian Hamiltonian dynamics solution for the global approach (see Eq.\,\eqref{eq:nhg}) is compared with the longest-lived state in panel ($b$). The parameters used in our simulations are:
    $\epsilon_c = \epsilon_h$, $\alpha_c = 0.2\epsilon_h$, $\alpha_h = 0.05\epsilon_h$, $T_c = 0.1\epsilon_h$, $\omega_c = 10\epsilon_h$, $g = 0.8\epsilon_h$. We tune the temperature $ T_h $ over the range $ T_h \in [0.15,1.50] \epsilon_h $ (see legend). 
        }
\end{figure}

Here, we show how both non-Hermitian Hamiltonian and Lindblad dynamics reach long-time states. We consider as initial state the thermal state at temperature $T_h$.
\begin{figure}[h] 
    \centering
    \includegraphics[scale=0.34]{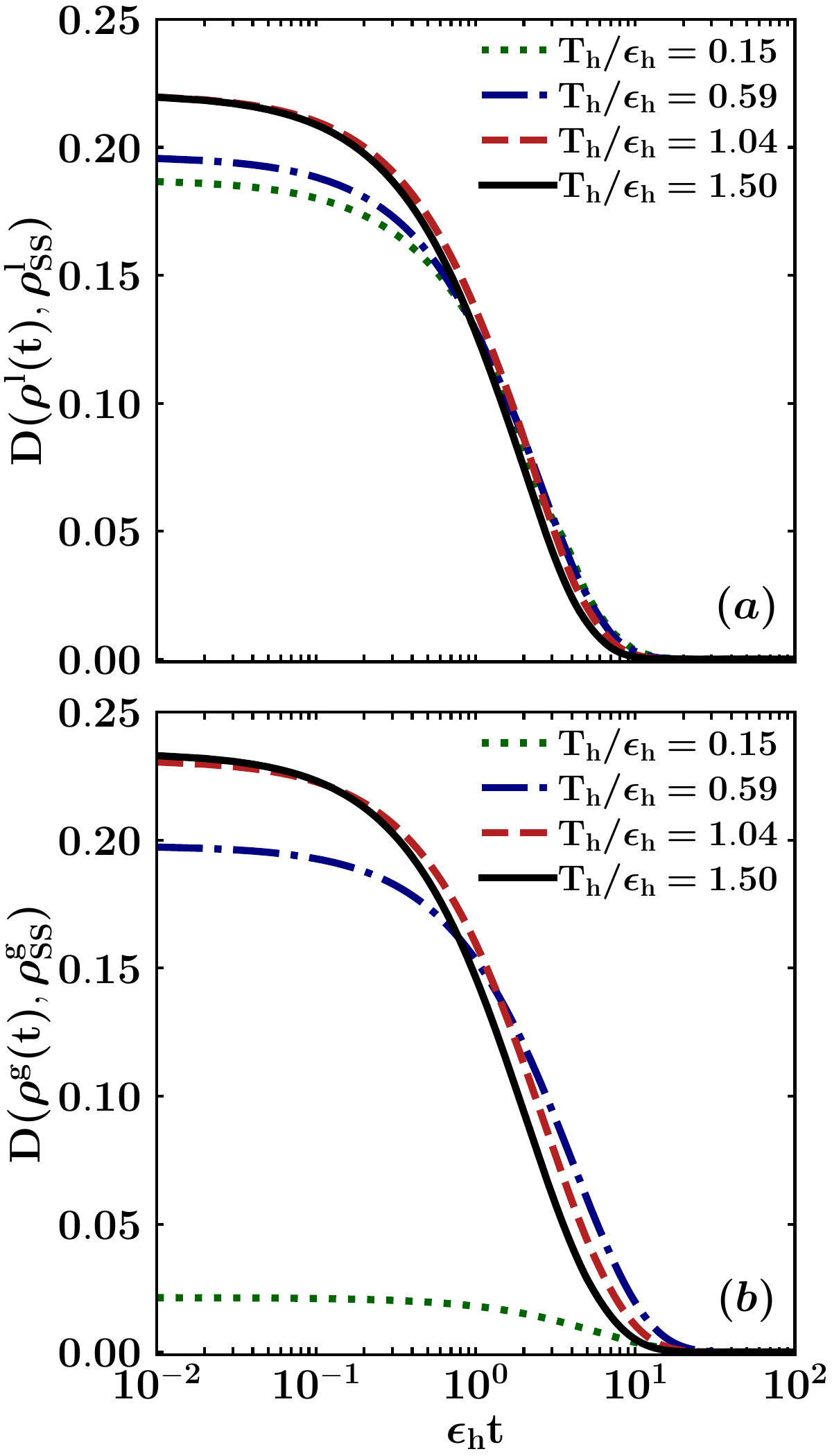} \caption{\label{fig:lind_lg_SS}
    Relaxation to the steady state in Lindblad dynamics, taking as the initial state the thermal state at temperature $T_h$ (temperature of the hot bath).
    Trace distance in Eq.\,\eqref{eq:trdist} as a function of dimensionless time for four values of $T_h$. The Lindblad evolved state $\rho_L(t)$ for the local approach (see Eq.\,\eqref{eq:dissipl}) is compared with the stationary state in panel ($a$). The same Lindblad dynamics solution for the global approach (see Eq.\,\eqref{eq:dissipg}) is compared with the stationary state in panel ($b$).
    The parameters used in our simulations are: 
    $\epsilon_c = \epsilon_h$, $\alpha_c = 0.2\epsilon_h$, $\alpha_h = 0.05\epsilon_h$, $T_c = 0.1\epsilon_h$, $\omega_c = 10\epsilon_h$, $g = 0.8\epsilon_h$. We tune the temperature $ T_h $ over the range $ T_h \in [0.15,1.50] \epsilon_h $ (see legend). 
        }
\end{figure}
Lindblad evolution in Fig.\,\ref{fig:lind_lg_SS} reaches stationary behavior for both the local and global approaches and displays a clear temperature dependence, which is more pronounced in the global case. In contrast, the non-Hermitian Hamiltonian dynamics shows very similar behavior in the local and global approaches: the local case is only slightly closer to the longest-lived state [Fig.\,\ref{fig:NH_lg_SS} (a)], whereas the global case remains farther from it over the same time window [Fig.\,\ref{fig:NH_lg_SS} (b)].

\section{Exceptional point in a microscopically valid local regime}
\label{app:ep_local_validity}

\begin{figure*}[ht]
    \begin{center}
    \includegraphics[width=.6\linewidth]{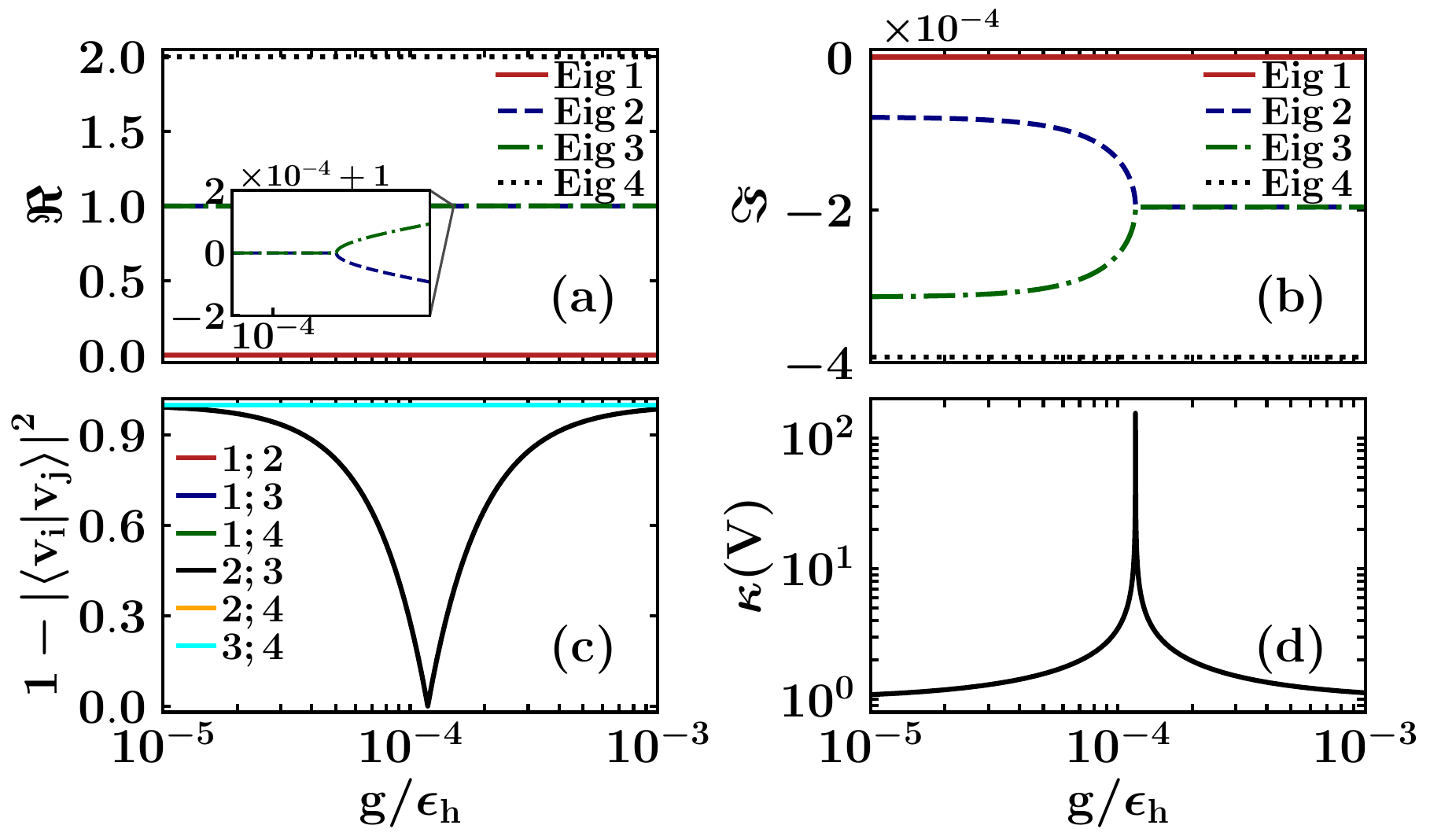}
    \caption{\label{fig:excpoints_app}
    Exceptional point in the non-Hermitian Hamiltonian derived from the local master equation.
    We report the coalescence of eigenvalues and eigenvectors 2 and 3. This is shown by examining the real part in panel ($a$) and the imaginary part in panel ($b$) of the eigenvalues as functions of $g$ (in units of $\epsilon_h$) with a zoomed-in view in panel ($a$). Non-orthogonality of eigenvectors 2 and 3 is computed by evaluating $1 - |\langle v_i | v_j \rangle|^2$ for each pair of right eigenvectors $i,j$ as a function of $g$ (in units of $\epsilon_h$) and shown in panel ($c$). Divergence of the condition number of the eigenvector matrix $\kappa(V)$ is shown in panel ($d$) as a function of the coupling $g$ (in units of $\epsilon_h$). The parameters used in our simulations are:
    $\epsilon_c = \epsilon_h$, $\alpha_c = 2\times10^{-4}\epsilon_h$, $\alpha_h = 5\times10^{-5}\epsilon_h$, $T_c = 0.1\epsilon_h$, $T_h = \epsilon_h$, $\omega_c = 20\epsilon_h$. We tune the coupling $g$ over the range $g \in [10^{-5},10^{-3}]\epsilon_h$ and find $\mathrm{EP}$ at $g \approx 1.2\times 10^{-4}\epsilon_h$.}
    \end{center}
\end{figure*}

In this appendix, we report the same four-panel analysis of the non-Hermitian Hamiltonian shown in Fig.\,\ref{fig:excpoints}, but for a parameter set chosen in the regime discussed in Ref.\,\cite{cattaneo2019local}, where the local approach is shown to be valid also when derived from an underlying microscopic Hamiltonian. For the precise parameter values, refer to the caption of Fig.\,\ref{fig:excpoints_app}. Within this regime we still observe an exceptional point, occurring at a coupling value $g \approx 10^{-4}\epsilon_h$, i.e., in a parameter region where the local approach is expected to be reliable according to Ref.\,\cite{cattaneo2019local}.

This additional result complements the discussion in the main text. There, we used a different parameter set, for which the microscopic validity of the local Lindblad master equation is not generally ensured. Nevertheless, the corresponding Markovian channels remain physically representable as completely positive trace-preserving maps, and therefore admit a unitary dilation by the Stinespring theorem.

Our focus is thus on the fundamental non-Hermitian features of Markovian master equations, while the present appendix shows that the exceptional-point phenomenology also persists in a regime of established microscopic validity for the local approach.
\section{Non-normality measure for a different parameter configuration}\label{app:nonnormality_max}
\begin{figure}[H]
    \centering
    \includegraphics[scale=0.29]{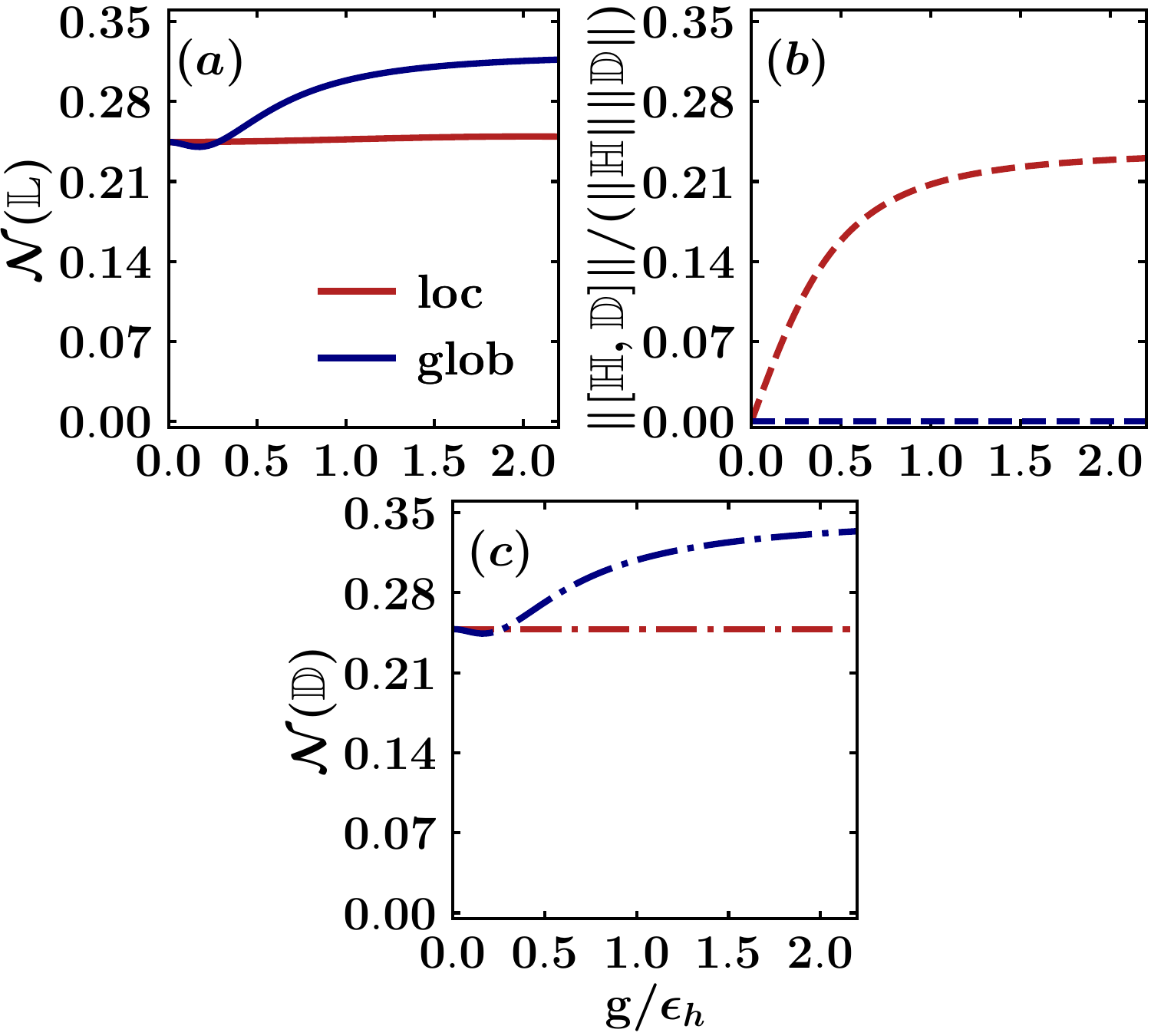}
    \caption{(a) Non-normality of the local (red) and global (blue) Lindbladians, defined in Eq.\,\eqref{eq:nonnormality}, shown as a function of the coupling $g$ (in units of $\epsilon_h$). (b) $[\mathbb{H},\mathbb{D}]$ contribution, $\left\Vert [\mathbb{H},\mathbb{D}] \right\Vert / (\left\Vert \mathbb{H} \right\Vert \,\left\Vert \mathbb{D} \right\Vert)$, as a function of $g/\epsilon_h$. (c) Non-normality of the dissipative part of the Lindbladian, $\mathbb{D}$, as a function of $g/\epsilon_h$. The parameters used in our simulations are: $\epsilon_c = 0.5\epsilon_h$, $\alpha_c = 2\epsilon_h$, $\alpha_h = 2\epsilon_h$, $T_c = 0.05\epsilon_h$, $T_h=0.2\epsilon_h$, $\omega_c = 10\epsilon_h$. We tune the coupling $g$ over the range $[0.0,2.2] \epsilon_h$. 
    This provides additional numerical results complementing Fig.\,\ref{fig:nonnorm}.}
    \label{fig:nonnorm_app}
\end{figure}
In Fig.\,\ref{fig:nonnorm_app} we report the Lindbladian non-normality $\mathcal{N}(\mathbb{L})$, the contribution arising from the commutator $[\mathbb{H},\mathbb{D}]$, and the dissipator non-normality $\mathcal{N}(\mathbb{D})$ for a different set of parameters.
These parameters are chosen by fixing the energy unit to $\epsilon_h$ and by maximizing the Lindbladian non-normality.

In Fig.\,\ref{fig:nonnorm_app}(a), the behavior of the local and global approaches is reversed with respect to the case analyzed in the main text (see Fig.\,\ref{fig:nonnorm}(a)). Also in this case, $\mathcal{N}(\mathbb{D})$ is constant for the local approach and increases for the global one, while the $[\mathbb{H},\mathbb{D}]$ contribution vanishes in the global approach and increases with $g$ in the local approach. 

Overall, Fig.\,\ref{fig:nonnorm_app} suggests that, even in this configuration, the behavior of $\mathcal{N}(\mathbb{L})$ is primarily determined by the dissipator contribution.


\nocite{apsrev42Control}
\bibliographystyle{apsrev4-2}
\bibliography{revtex-control,biblio}

\end{document}